\documentclass[journal,11pt,draftclsnofoot,onecolumn]{IEEEtran}

\usepackage{latexsym}
\usepackage{amsfonts}
\usepackage{graphicx} 

\usepackage{amssymb}
\usepackage{stmaryrd}
\usepackage{natbib}
\usepackage{enumerate}
\usepackage{amsmath}

\newcommand{\dfn}{\triangleq}
\newcommand{\bftheta}{{\mbox{\boldmath $\vartheta$}}}

\def\BibTeX{{\rm B\kern-.05em{\sc i\kern-.025em b}\kern-.08em
    T\kern-.1667em\lower.7ex\hbox{E}\kern-.125emX}}

\begin{document}

%
\title{Generalized Rejection Sampling Schemes and Applications in Signal Processing}
\author{Luca Martino and Joaqu\'{\i}n M\'{\i}guez\\
Department of Signal Theory and Communications, Universidad Carlos III de Madrid.\\
Avenida de la Universidad 30, 28911 Legan\'es, Madrid, Spain.\\
E-mail: {\tt luca@tsc.uc3m.es, joaquin.miguez@uc3m.es}}
\maketitle

\vspace{-1.5cm}
\begin{abstract}
Bayesian methods and their implementations by means of sophisticated Monte Carlo techniques, such as Markov chain Monte Carlo (MCMC) and particle filters, have become very popular in signal processing over the last years. However, in many problems of practical interest these techniques demand procedures for sampling from probability distributions with non-standard forms, hence we are often brought back to the consideration of fundamental simulation algorithms, such as rejection sampling (RS). Unfortunately, the use of RS techniques demands the calculation of tight upper bounds for the ratio of the target probability density function (pdf) over the proposal density from which candidate samples are drawn. Except for the class of log-concave target pdf's, for which an efficient algorithm exists, there are no general methods to analytically determine this bound, which has to be derived from scratch for each specific case. In this paper, we introduce new schemes for (a) obtaining upper bounds for likelihood functions and (b) adaptively computing proposal densities that approximate the target pdf closely. The former class of methods provides the tools to easily sample from {\it a posteriori} probability distributions (that appear very often in signal processing problems) by drawing candidates from the prior distribution. However, they are even more useful when they are exploited to derive the generalized adaptive RS (GARS) algorithm introduced in the second part of the paper. The proposed GARS method yields a sequence of proposal densities that converge towards the target pdf and enable a very efficient sampling of a broad class of probability distributions, possibly with multiple modes and non-standard forms. We provide some simple numerical examples to illustrate the use of the proposed techniques, including an example of target localization using range measurements, often encountered in sensor network applications.
\end{abstract}
\IEEEpeerreviewmaketitle

\section{Introduction}
Bayesian methods have become very popular in signal processing during the past decades and, with them, there has been a surge of interest in the Monte Carlo techniques that are often necessary for the implementation of optimal {\em a posteriori} estimators \cite{Fitzgerald01,Doucet01b,Robert04,Liu04b}. Indeed, Monte Carlo statistical methods are powerful tools for numerical inference and optimization \cite{Robert04}. Currently, there exist several classes of MC techniques, including the popular Markov Chain Monte Carlo (MCMC) \cite{Fitzgerald01,Larocque02} and particle filtering \cite{Djuric03,Doucet01b} families of algorithms, which enjoy numerous applications. However, in many problems of practical interest these techniques demand procedures for sampling from probability distributions with non-standard forms, hence we are often brought back to the consideration of fundamental simulation algorithms, such as importance sampling \cite{DeGroot02}, inversion procedures \cite{Robert04} and the accept/reject method, also known as {\em rejection sampling} (RS).

The RS approach \cite[Chapter 2]{Robert04} is a classical Monte Carlo technique for ``universal sampling''. It can be used to generate samples from a target probability density function (pdf) by drawing from a possibly simpler proposal density. The sample is either accepted or rejected by an adequate test of the ratio of the two pdf's, and it can be proved that accepted samples are actually distributed according to the target density. RS can be applied as a tool by itself, in problems where the goal is to approximate integrals with respect to (w.r.t.) the pdf of interest, but more often it is a useful building block for more sophisticated Monte Carlo procedures \cite{Gilks92,Gilks94,Kunsch05}. An important limitation of RS methods is the need to analytically establish a bound for the ratio of the target and proposal densities, since there is a lack of general methods for the computation of exact bounds. 

One exception is the so-called {\em adaptive rejection sampling} (ARS) method \cite{Gilks92,Gilks92derfree,Robert04} which, given a target density, provides a procedure to obtain both a suitable proposal pdf (easy to draw from) and the upper bound for the ratio of the target density over this proposal. Unfortunately, this procedure is only valid when the target pdf is strictly log-concave, which is not the case in most practical cases. 
Although an extension has been proposed \cite{Hoermann95,Evans98} that enables the application of the ARS algorithm with $T$-concave distributions (where $T$ is a monotonically increasing function, not necessarily the logarithm), it does not address the main limitations of the original method (e.g., the impossibility to draw from multimodal distributions) and is hard to apply, due to the difficulty to find adequate $T$ transformations other than the logarithm. Another algorithm, called {\em adaptive rejection metropolis sampling} (ARMS) \cite{Gilks95}, is an attempt to extend the ARS to multimodal densities by adding Metropolis-Hastings steps. However, the use of an MCMC procedure has two important consequences. First, the resulting samples are correlated (unlike in the original ARS method), and, second, for multimodal distributions the Markov Chain often tends to get trapped in a single mode.  

In this paper we propose general procedures to apply RS when the target pdf is the posterior density of a signal of interest (SoI) given a collection of observations. Unlike the ARS technique, our methods can handle target pdf's with several modes (hence non-log-concave) and, unlike the ARMS algorithm, they do not involve MCMC steps. Hence, the resulting samples are independent and come exactly from the target pdf. 

We first tackle the problem of computing an upper bound for the likelihood of the SoI given fixed observations. The proposed solutions, that include both closed-form bounds and iterative procedures, are useful when we draw the candidate samples from the prior pdf. 

In this second part of the paper, we extend our approach to devise a generalization of the ARS method that can be applied to a broad class of pdf's, possibly multimodal. The generalized algorithm yields an efficient proposal density, tailored to the target density, that can attain a much better acceptance rate than the prior distribution. We remark that accepted samples from the target pdf are independent and identically distributed (i.i.d). 

The remaining of the paper is organized as follows. We formally describe the signal model in Section \ref{Modelandstat}. Some useful definitions and basic assumptions are introduced in Section \ref{sectPr}. In Section \ref{sectSL2}, we propose a general procedure to compute upper bounds for a large family of likelihood functions. 
The ARS method is briefly reviewed in Section \ref{secARS}, while the main contribution of the paper, the generalization of the ARS algorithm, is introduced in Section \ref{secMARS}. Section \ref{sectExample} is devoted to simple numerical examples and we conclude with a brief summary in Section \ref{sConclusions}.

\section{Model and Problem Statement}
\label{Modelandstat}

\subsection{Notation}
Scalar magnitudes are denoted using regular face letters, e.g., $x$, $X$, while vectors are displayed
as bold-face letters, e.g., $\textbf{x}$, $\textbf{X}$. We indicate random variates with upper-case letters, e.g.,  $X$, $\textbf{X}$, while we use lower-case letters to denote the corresponding realizations, e.g., $x$, $\textbf{x}$.  We use letter $p$ to denote the true probability density function (pdf) of a random variable or vector. This is an argument-wise notation, common in Bayesian analysis. For two random variables $X$ and $Y$, $p(x)$ is the true pdf of $X$ and $p(y)$ is the true pdf of $Y$, possibly different. The conditional pdf of $X$ given $Y=y$ is written $p(x|y)$. Sets are denoted with calligraphic upper-case letters, e.g., $\mathcal{R}$. 

\subsection{Signal Model}
\label{secSignalModel}
Many problems in science and engineering involve the estimation of an unobserved SoI, $\textbf{x}\in \mathbb{R}^{m}$, from a sequence of related observations. We assume an arbitrary prior probability density function (pdf) for the SoI, $\textbf{X}\sim p(\textbf{x})$, and consider $n$ scalar random observations, $Y_{i}\in \mathbb{R}$, $i=1,\ldots,n$, which are obtained through nonlinear transformations of the signal \textbf{X} contaminated with additive noise. Formally, we write
\begin{equation}
\label{sistemasensores}
Y_{1}=g_{1}(\textbf{X})+\Theta_{1}, \ldots, Y_{n}=g_{n}(\textbf{X})+\Theta_{n}
\end{equation}   
where $\textbf{Y}=[Y_{1},\ldots,Y_{n}]^{\top}\in\mathbb{R}^{n}$ is the random observation vector, $g_{i}:\mathbb{R}^{m}\rightarrow\mathbb{R}, \ \ i=1,\ldots,n$, are nonlinearities and $\Theta_{i}$ are independent noise variables, possibly with different distributions for each $i$. We write $\textbf{y}=[y_{1},\ldots,y_{n}]^{\top}\in\mathbb{R}^{n}$ for the vector of available observations, i.e., a realization of $\textbf{Y}$. 

We assume exponential-type noise pdf's, of the form  
\begin{eqnarray}
\label{noise}
\Theta_{i}\sim	p(\vartheta_{i}) = k_{i} \exp\left\{-\bar{V}_{i}(\vartheta_{i})\right\}, 
\end{eqnarray}
where $k_{i}>0$ is real constant and $\bar{V}_i(\vartheta_{i})$ is a function, subsequently referred to as marginal potential, with the following properties: 
\begin{enumerate} 
\renewcommand{\labelenumi}{\Roman{enumi}.}
  \item[(P1)] It is real and non negative, i.e., $\bar{V}_{i}: \mathbb{R}\rightarrow [0,+\infty)$.  
	\item[(P2)] It is increasing ($\frac{d\bar{V}_{i}}{d\vartheta_{i}}>0$) for  $\vartheta_{i}>0$ and decreasing ($\frac{d\bar{V}_{i}}{d\vartheta_{i}}<0$) for  $\vartheta_{i}<0$.  
\end{enumerate}
These conditions imply that $\bar{V}_{i}(\vartheta_{i})$ has a unique minimum at $\vartheta_{i}^{*}=0$ and, as a consequence $p(\vartheta_{i})$ has only one maximum (mode) at $\vartheta^{*}_{i}=0$. Since the noise variables are independent, the joint pdf $p(\vartheta_{1},\vartheta_{2},\ldots,\vartheta_{n})=\prod_{i=1}^{n}p(\vartheta_{n})$ is easy to construct and we can define a joint potential function $V^{(n)}: \mathbb{R}^{n}\rightarrow [0,+\infty)$ as
\begin{equation}
\label{jointpotential}
	V^{(n)}(\vartheta_{1},\ldots,\vartheta_{n})\dfn-\log \left[p(\vartheta_{1},\ldots,\vartheta_{n})\right]
	=-\sum_{i=1}^{n}\log [p(\vartheta_{n})].
\end{equation}
Substituting (\ref{noise}) into (\ref{jointpotential}) yields
\begin{equation}
\label{genPot2}
V^{(n)}(\vartheta_{1},\ldots,\vartheta_{n})=c_{n}+\sum^{n}_{i=1}\bar{V}_{i}(\vartheta_{i})
\end{equation}
where $c_{n}=-\sum^{n}_{i=1}\log{k_{i}}$ is a constant. In subsequent sections we will be interested in a particular class of joint potential functions denoted as
\begin{eqnarray}
V_{l}^{(n)}(\vartheta_{1},\ldots,\vartheta_{n})=\sum^{n}_{i=1}\left|\vartheta_{i}\right|^{l},	 \ \ \ \ 0<l<+\infty,
\end{eqnarray} 
where the subscript $l$ identifies the specific member of the class. In particular, the function obtained for $l=2$,  
$V_{2}^{(n)}(\vartheta_{1},\ldots,\vartheta_{n})=\sum^{n}_{i=1}\left|\vartheta_{i}\right|^{2}$ is termed quadratic potential.

Let $\textbf{g}=[g_{1},\ldots,g_{n}]^{\top}$ be the vector-valued  nonlinearity defined as $\textbf{g}(\textbf{x})\dfn[g_{1}(\textbf{x}),\ldots,g_{n}(\textbf{x})]^{\top}$.
The scalar observations are conditionally independent given a realization of the SoI, $\textbf{X}=\textbf{x}$, hence the \textsl{likelihood function} $\ell(\textbf{x};\textbf{y},\textbf{g})\dfn p(\textbf{y}|\textbf{x})$, can be factorized as
\begin{eqnarray}
\label{like}
\ell(\textbf{x};\textbf{y},\textbf{g})=\prod^{n}_{i=1} p(y_{i}|\textbf{x}),
\end{eqnarray}
where $p(y_{i}|\textbf{x})=k_{i}\exp\left\{-\bar{V}_{i}(y_{i}-g_{i}(\textbf{x}))\right\}$. The likelihood in (\ref{like}) induces a system potential function $V(\textbf{x};\textbf{y},\textbf{g}): \mathbb{R}^{m}\rightarrow [0,+\infty)$, defined as 
\begin{eqnarray}
\label{SystemPot}
V(\textbf{x};\textbf{y},\textbf{g})\dfn-\log[\ell(\textbf{x};\textbf{y},\textbf{g})]=-\sum^{n}_{i=1}\log[p(y_{i}|\textbf{x})],
\end{eqnarray}
that depends on \textbf{x}, the observations \textbf{y}, and the function \textbf{g}. Using (\ref{genPot2}) and (\ref{SystemPot}), we can write the system potential in terms of the joint potential,
\begin{equation}
\label{genPot}
\small
V(\textbf{x};\textbf{y},\textbf{g})=c_{n}+\sum^{n}_{i=1}\bar{V}_{i}(y_{i}-g_{i}(\textbf{x}))= V^{(n)}(y_{1}-g_{1}(\textbf{x}),\ldots,y_{n}-g_{n}(\textbf{x})).
\end{equation}

\subsection{Rejection Sampling}
Assume that we wish to approximate, by sampling, some integral of the form $I(f)=\int_{\mathbb{R}} f(\textbf{x})p(\textbf{x}|\textbf{y})d\textbf{x}$, where $f$ is some measurable function of $\textbf{x}$ and $p(\textbf{x}|\textbf{y}) \propto p(\textbf{x})\ell(\textbf{x};\textbf{y},\textbf{g})$ is the posterior pdf of the SoI given the observations. Unfortunately, it may not be possible in general to draw directly from $p(\textbf{x}|\textbf{y})$, so we need to apply simulation techniques to generate adequate samples. One appealing possibility is to perform RS using the prior, $p(\textbf{x})$, as a proposal function. In such case, let $\gamma$ be a lower bound for the system potential, $\gamma \leq V(\textbf{x};\textbf{y},\textbf{g})$, so that $L\dfn\exp\{-\gamma\}$ is an upper bound for the likelihood, $\ell(\textbf{x};\textbf{y},\textbf{g}) \leq L$. We can generate $N$ samples according to the standard RS algorithm.
\begin{enumerate}
\item Set $i=1$.
\item Draw samples $\textbf{x}'$ from $p(\textbf{x})$ and $u'$ from $U(0,1)$, where $U(0,1)$ is the uniform pdf in $[0,1]$.
\item If $\frac{p(\textbf{x}'|\textbf{y})}{L p(\textbf{x}')} \propto \frac{\ell(\textbf{x}';\textbf{y},\textbf{g})}{L}> u'$ 
then $\textbf{x}^{(i)}=\textbf{x}'$, else discard $\textbf{x}'$ and go back to step 2.
\item Set $i=i+1$. If $i>N$ then stop, else go back to step 2.  
\end{enumerate}
Then, $I(f)$ can be approximated as $I(f)\approx\hat I(f)=\frac{1}{N}\sum_{i=1}^N f(\textbf{x}^{(i)})$. The fundamental figure of merit of a rejection sampler is the acceptance rate, i.e., the mean number of accepted samples over the total number of proposed candidates.

In Section \ref{sectSL2}, we address the problem of analytically calculating the bound $L=\exp\{-\gamma\}$. Note that, since the $\log$ function is monotonous, it is equivalent to maximize $\ell$ w.r.t. \textbf{x} and to minimize the system potential $V$ also w.r.t. \textbf{x}.  As a consequence, we may focus on the calculation of a lower bound $\gamma$ for $V(\textbf{x};\textbf{y},\textbf{g})$. Note that this problem is far from trivial. Even for very simple marginal potentials, $\bar{V}_{i}$, $i=1,...,n$, the system potential can be highly multimodal w.r.t. \textbf{x}. See the example in the Section \ref{Example1} for an illustration.

\section{Definitions and Assumptions} 
\label{sectPr}
Hereafter, we restrict our attention to the case of a scalar SoI, $x\in\mathbb{R}$. This is done for the sake of clarity, since dealing with the general case $\textbf{x}\in \mathbb{R}^{m}$ requires additional definitions and notations. The techniques to be described in Sections \ref{sectSL2}-\ref{secMARS} can be extended to the general case, although this extension is not trivial. The example in Section \ref{Example4} illustrates how the proposal methodology is also useful in higher dimensional spaces, though.

For a given vector of observations $\textbf{Y}=\textbf{y}$, we define the set of \textit{simple estimates} of the SoI as
\begin{equation}
\mathcal{X}\dfn \left\{ x_{i}\in \mathbb{R}: \ \ y_{i}=g_{i}(x_{i}) \ \ \mbox{for} \ i=1,\ldots,n  \right\}.
\end{equation}
Each equation $y_{i}=g_{i}(x_{i})$, in general, can yield zero, one or more simple estimates. We also introduce the maximum likelihood (ML) SoI estimator $\hat{x}$, as
\begin{equation}	
\hat{x}\in \arg\max\limits_{x\in\mathbb{R}}{\ell(x|\textbf{y},\textbf{g})}=\arg\min\limits_{x\in\mathbb{R}}{V(x;\textbf{y},\textbf{g})},
\end{equation} 
not necessarily unique.
 
Let us use $\mathcal{A}\subseteq \mathbb{R}$ to denote the support of the vector function $\textbf{g}$, i.e.,  $\textbf{g}:\mathcal{A}\subseteq \mathbb{R}\rightarrow \mathbb{R}^{n}$. We assume that there exists a partition $\{\mathcal{B}_{j}\}_{j=1}^{q}$ of $\mathcal{A}$ (i.e., $\mathcal{A}=\cup_{j=1}^{q} \mathcal{B}_{j}$ and $\mathcal{B}_{i}\cap \mathcal{B}_{j}=\emptyset$, $\forall i\neq j$) such that the subsets $\mathcal{B}_{j}$ are intervals in $\mathbb{R}$ and we can define functions $g_{i,j}: \mathcal{B}_{j}\rightarrow \mathbb{R}, \ \ \ j=1,\ldots,q$ and $i=1,\ldots,n$, as
\begin{equation}
g_{i,j}(x)\dfn g_{i}(x), \ \ \forall x\in \mathcal{B}_{j},	
\end{equation}
i.e., $g_{i,j}$ is the restriction of $g_{i}$ to the interval $\mathcal{B}_{j}$. We further assume that 
(a) every function $g_{i,j}$ is invertible in $\mathcal{B}_{j}$ and (b) every function $g_{i,j}$ is either convex in $\mathcal{B}_{j}$ or concave in $\mathcal{B}_{j}$. Assumptions (a) and (b) together mean that, for every $i$ and all $x \in \mathcal{B}_{j}$, the first derivative $\frac{d g_{i,j}}{dx}$ is either strictly positive or strictly negative and the second derivative $\frac{d^{2} g_{i,j}}{dx^{2}}$ is either non-negative or non-positive. As a consequence, there are exactly $n$ simple estimates (one per observation) in each subset of the partition, namely $x_{i,j}=g_{i,j}^{-1}(y_{i})$ for $i=1,\ldots,n$. We write the set of simple estimates in $\mathcal{B}_{j}$ as $\mathcal{X}_{j}=\{x_{1,j},\ldots,x_{n,j}\}$. Due to the additivity of the noise in (\ref{sistemasensores}), if $g_{i,j}$ is bounded there may be a non-negligible probability that  $Y_{i}>\max\limits_{x\in[\mathcal{B}_{j}]} g_{i,j}(x)$ (or $Y_{i}<\min\limits_{x\in[\mathcal{B}_{j}]} g_{i,j}(x)$), where $[\mathcal{B}_{j}]$ denotes the closure of set $\mathcal{B}_{j}$, hence $g_{i,j}^{-1}(y_{i})$ may not exist for some realization $Y_i=y_i$. In such case, we define $x_{i,j}= \arg\max\limits_{x\in[\mathcal{B}_{j}]} g_{i,j}(x)$ (or $x_{i,j}=\arg\min\limits_{x\in[\mathcal{B}_{j}]} g_{i,j}(x)$, respectively), and admit $x_{i,j}=+\infty$ (respectively, $x_{i,j}=-\infty$) as valid simple estimates. 

\section{Computation of upper bounds on the likelihood}
\label{sectSL2}
\subsection{Basic method}
\label{sectBM1}
Let $\textbf{y}$ be an arbitrary but fixed realization of the observation vector $\textbf{Y}$.
Our goal is to obtain an analytical method for the computation of a scalar $\gamma(\textbf{y})\in \mathbb{R}$ such that $\gamma(\textbf{y})\leq \inf\limits_{x\in \mathbb{R}}V(x;\textbf{y},\textbf{g})$. Hereafter, we omit the dependence on the observation vector and write simply $\gamma$. The main difficulty to carry out this calculation is the nonlinearity $\textbf{g}$, which renders the problem not directly tractable. To circumvent this obstacle, we split the problem into $q$ subproblems and address the computation of bounds for each set $\mathcal{B}_{j}$, $j=1,\ldots,q$, in the partition of $\mathcal{A}$. Within $\mathcal{B}_{j}$, we build adequate linear functions $\left\{r_{i,j}\right\}_{i=1}^{n}$ in order to replace the nonlinearities $\left\{g_{i,j}\right\}_{i=1}^{n}$. We require that, for every $r_{i,j}$, the inequalities  
\begin{equation}
\label{condR1}
	\left|y_{i}-r_{i,j}(x)\right| \leq \left|y_{i}-g_{i,j}(x)\right|, \ \mbox{and}  
\end{equation}
\begin{equation}
\label{condR2}
(y_{i}-r_{i,j}(x))(y_{i}-g_{i,j}(x))\geq 0  
\end{equation} 
hold jointly for all $i=1,\ldots,n$, and all $x\in \mathcal{I}_{j}\subset \mathcal{B}_{j}$, where $\mathcal{I}_{j}$ is any closed interval in $\mathcal{B}_{j}$ such that $\hat{x}_{j}\in \arg \min\limits_{x\in [\mathcal{B}_{j}]} V(x;\textbf{y},\textbf{g})$ (i.e., any ML estimator of the SoI $X$ restricted  to $\mathcal{B}_{j}$, possibly non unique) is contained in $\mathcal{I}_{j}$. The latter requirement can be fulfilled if we choose $\mathcal{I}_{j}\dfn [\min(\mathcal{X}_{j}),\max(\mathcal{X}_{j})]$ (see the Appendix for a proof). \\
If (\ref{condR1}) and (\ref{condR2}) hold, we can write 
\begin{equation}
\label{underbound}
	\bar{V}_{i}(y_{i}-r_{i,j}(x)) \leq \bar{V}_{i}(y_{i}-g_{i,j}(x)), \ \ \ \forall x\in \mathcal{I}_{j}, 
\end{equation} 
which follows easily from the properties (P1) and (P2) of the marginal potential functions $\bar{V}_{i}$ as described in Section \ref{secSignalModel}. Moreover, since  $V(x;\textbf{y},\textbf{g}_{j})=c_{n}+\sum_{i=1}^{n}\bar{V}_{i}(y_{i}-g_{i,j}(x))$ and $V(x;\textbf{y},\textbf{r}_{j})=c_{n}+\sum_{i=1}^{n}\bar{V}_{i}(y_{i}-r_{i,j}(x))$ (this function will be subsequently referred as the modified system potential) where $\textbf{g}_{j}(x)\dfn[g_{1,j}(x),\ldots,g_{n,j}(x)]$ and $\textbf{r}_{j}(x)\dfn[r_{1,j}(x),\ldots,r_{n,j}(x)]$, Eq. (\ref{underbound}) implies that 
$V(x;\textbf{y},\textbf{r}_{j})\leq V(x;\textbf{y},\textbf{g}_{j})$, $\forall x\in \mathcal{I}_{j}$, and, as a consequence, 
\begin{eqnarray}
\gamma_{j}=\inf\limits_{x\in \mathcal{I}_{j}} {V(x;\textbf{y},\textbf{r}_{j})}\leq \inf\limits_{x\in \mathcal{I}_{j}}{V(x;\textbf{y},\textbf{g}_{j})}=\inf\limits_{x\in \mathcal{B}_{j}}{V(x;\textbf{y},\textbf{g})}.
\end{eqnarray}   
Therefore, it is possible to find a lower bound in $\mathcal{B}_{j}$ for the system potential $V(x;\textbf{y},\textbf{g}_{j})$, denoted $\gamma_{j}$, by minimizing the modified potential $V(x;\textbf{y},\textbf{r}_{j})$ in $\mathcal{I}_{j}$.

All that remains is to actually build the linearities $\left\{r_{i,j}\right\}_{i=1}^{n}$. This construction is straightforward and can be described graphically by splitting the problem into two cases. Case 1 corresponds to nonlinearities $g_{i,j}$ such that $\frac{d g_{i,j}(x)}{d x}\times \frac{d^{2}g_{i,j}(x)}{d x^{2}}\geq 0$ (i.e., $g_{i,j}$ is either increasing and convex or decreasing and concave), while case 2 corresponds to functions that comply with $\frac{d g_{i,j}(x)}{d x}\times \frac{d^{2}g_{i,j}(x)}{d x^{2}}\leq 0$ (i.e., $g_{i,j}$ is either increasing and concave or decreasing and convex), when $x\in \mathcal{B}_{j}$. 

\begin{figure}[htb]
\centerline{
 		\includegraphics[width=12cm]{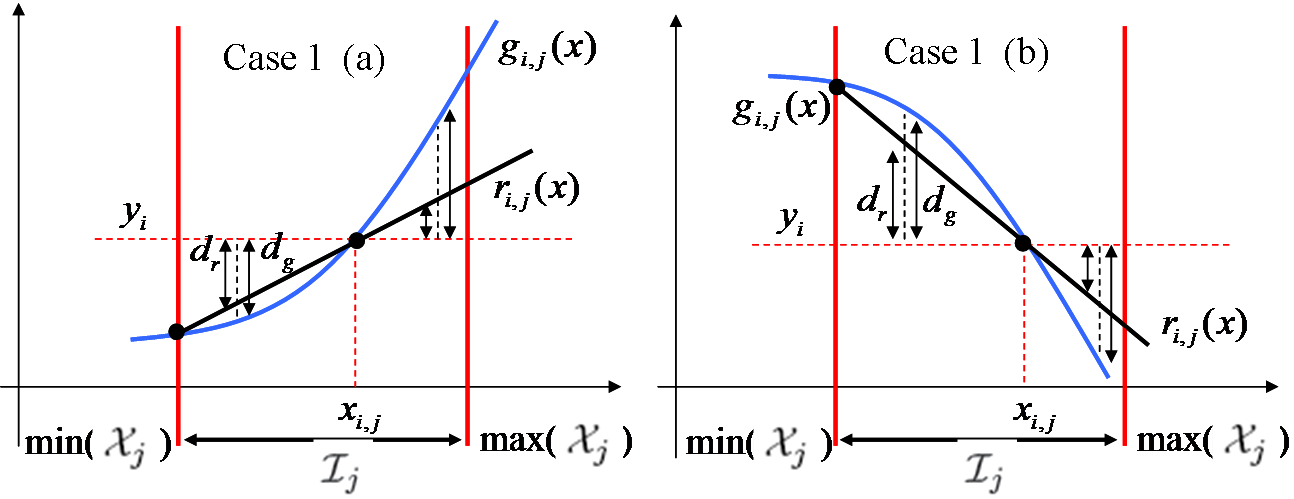} 
 		}
\centerline{ 
 		\includegraphics[width=12cm]{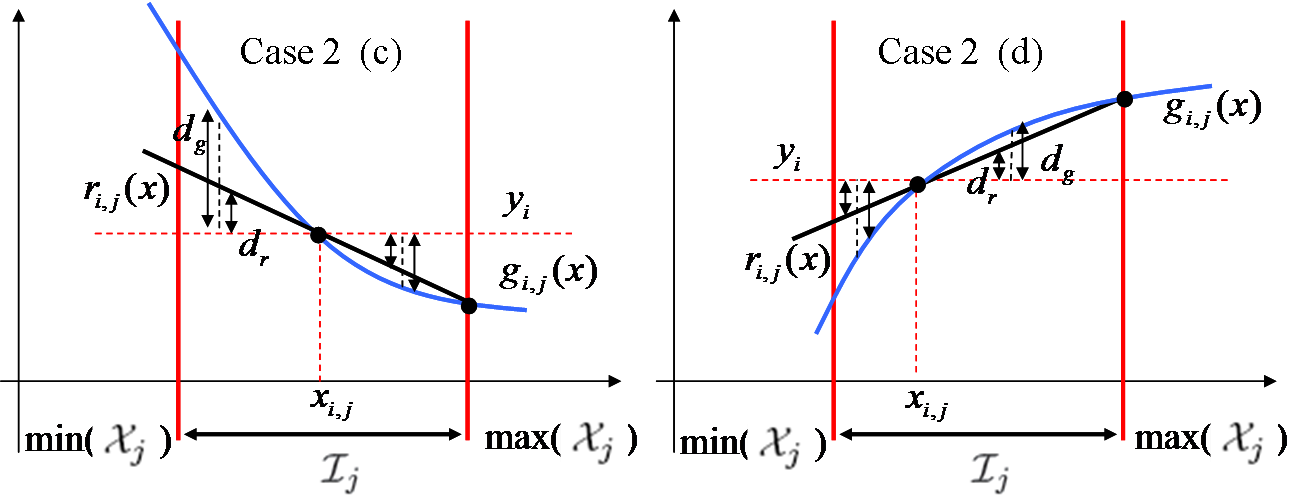}
 		}
\caption{Construction of the auxiliary linearities $\left\{r_{i,j}\right\}_{i=1}^{n}$. We indicate $d_{r}=\left|y_{i}-r_{i,j}(x)\right|$ and $d_{g}=\left|y_{i}-g_{i,j}(x)\right|$, respectively. It is apparent that $d_{r}\leq d_{g}$ and $r_{i,j}(x)$ and $g_{i,j}(x)$ are either simultaneously greater than (or equal to) $y_{i}$, or simultaneously lesser than (or equal to) $y_{i}$, for all $x\in \mathcal{I}_{j}$. Hence, the inequalities (\ref{condR1}) and (\ref{condR2}) are satisfied $\forall x\in \mathcal{I}_{j}$. (a) Function $g_{i,j}$ is increasing and convex (case 1). (b) Function $g_{i,j}$ is decreasing and concave (case 1). (c) Function $g_{i,j}$ is decreasing and convex (case 2). (d) Function $g_{i,j}$ is increasing and concave (case 2).}
\label{figcase1}
\end{figure}

Figure \ref{figcase1} (a)-(b) depicts the construction of $r_{i,j}$ in case 1. We choose a linear function $r_{i,j}$
that connects the point $(\min{(\mathcal{X}_{j})},g(\min{(\mathcal{X}_{j})}))$ and the point corresponding to the simple estimate, $(x_{i,j},g(x_{i,j}))$. In the figure, $d_{r}$ and $d_{g}$ denote the distances $\left|y_{i}-r_{i,j}(x)\right|$ and $\left|y_{i}-g_{i,j}(x)\right|$, respectively. It is apparent that $d_{r}\leq d_{g}$ for all $x\in \mathcal{I}_{j}$, hence inequality (\ref{condR1}) is granted. Inequality (\ref{condR2}) also holds for all $x\in \mathcal{I}_{j}$, since $r_{i,j}(x)$ and $g_{i,j}(x)$ are either simultaneously greater than (or equal to) $y_{i}$, or simultaneously lesser than (or equal to) $y_{i}$. 

Figure \ref{figcase1} (c)-(d) depicts the construction of $r_{i,j}$ in case 2. We choose a linear function $r_{i,j}$ that connects the point $(\max{(\mathcal{X}_{j})},g(\max{(\mathcal{X}_{j})}))$ and the point corresponding to the simple estimate, $(x_{i,j},g(x_{i,j}))$. Again, $d_{r}$ and $d_{g}$ denote the distances $\left|y_{i}-r_{i,j}(x)\right|$ and  $\left|y_{i}-g_{i,j}(x)\right|$, respectively. It is apparent from the two plots that inequalities (\ref{condR1}) and (\ref{condR2}) hold for all $x\in \mathcal{I}_{j}$.

A special subcase of 1 (respectively, of 2) occurs when $x_{i}=\min{(\mathcal{X}_{j})}$ (respectively, $x_{i,j}=\max{(\mathcal{X}_{j})}$). Then, $r_{i,j}(x)$ is the tangent to $g_{i,j}(x)$ at $x_{i,j}$. If $x_{i,j}=\pm\infty$ then $r_{i,j}(x)$ is a horizontal asymptote of $g_{i,j}(x)$.

It is often possible to find $\gamma_{j}=\inf\limits_{x\in \mathcal{I}_{j}} V(x;\textbf{y},\textbf{r}_{j})\leq \inf\limits_{x\in \mathcal{I}_{j}}V(x;\textbf{y},\textbf{g}_{j})$ in closed-form. If we choose $\gamma=\min\limits_{j}\gamma_{j}$, then $\gamma\leq \inf\limits_{x\in \mathbb{R}} V(x,\textbf{y},\textbf{g})$ is a global lower bound of the system potential. Table \ref{tabla1} shows an outline of the proposed method, that will be subsequently referred to as bounding method 1 (BM1) for conciseness.

\begin{table}[!hbt]
\caption{Bounding Method 1.}
\label{tabla1}
\begin{center}
\begin{tabular}{||l||}
\hline
\hline
1. Find a partition $\left\{ \mathcal{B}_{j} \right\}_{j=1}^{q}$ of the space of the SoI. \\
2. Compute the simple estimates $\mathcal{X}_{j}=\left\{x_{1,j},\ldots,x_{n,j}\right\}$ for each $\mathcal{B}_{j}$. \\
3. Calculate $\mathcal{I}_{j}\dfn[\min(\mathcal{X}_{j}),\max(\mathcal{X}_{j})]$ and build $r_{i,j}(x)$, for $x\in \mathcal{I}_{j}$ and $i=1,\ldots,n$. \\ 
4. Replace $\textbf{g}_{j}(x)$ with $\textbf{r}_{j}(x)$, and minimize $V(x;\textbf{y},\textbf{r}_{j})$ to find the lower bound $\gamma_{j}$.\\
5. Find $\gamma=\min\limits_{j} \gamma_{j}$. \\
\hline
\hline
\end{tabular}
\end{center}
\end{table}
 
\subsection{Iterative Implementation}
\label{improve}
The quality of the bound $\gamma_{j}$ depends, to a large extent, on the length of the interval $\mathcal{I}_{j}$, denoted $|\mathcal{I}_j|$. This is clear if we think of $r_{i,j}(x)$ as a linear approximation on $\mathcal{I}_{j}$ of the nonlinearity $g_{i,j}(x)$. Since we have assumed $g_{i,j}(x)$ is continuous and bounded in $\mathcal{I}_{j}$, the procedure to build $r_{i,j}(x)$ in BM1 implies that
\begin{equation}
	\lim_{\left|\mathcal{I}_{j}\right|\rightarrow 0} \left|g_{i,j}(x)-r_{i,j}(x)\right|\leq \lim_{\left|\mathcal{I}_{j}\right|\rightarrow 0} |\sup_{x\in \mathcal{I}_{j}}g_{i,j}(x)-\inf_{x\in \mathcal{I}_{j}}g_{i,j}(x)|= 0,
\end{equation}
for all $x\in \mathcal{I}_{j}$. Therefore, if we consider intervals $\mathcal{I}_{j}$ which are shorter and shorter, then the modified potential function $V(x;\textbf{y},\textbf{r}_j)$ will be closer and closer to the true potential function $V(x;\textbf{y},\textbf{g}_j)$, and hence the bound $\gamma_j\leq V(x;\textbf{y},\textbf{r}_j)\leq V(x;\textbf{y},\textbf{g}_j)$ will be tighter. 

The latter observation suggests a procedure to improve the bound $\gamma_j$ for a given interval $\mathcal{I}_j$. Indeed, let us subdivide $\mathcal{I}_j$ into $k$ subintervals denoted $\mathcal{I}_{v,v+1}\dfn[s_{v},s_{v+1}]$ where $v=1,\ldots,k$ and $s_{v},s_{v+1} \in \mathcal{I}_{j}$. We refer to the elements in the collection $\mathcal{S}_{j,k}=\{s_1,\ldots, s_{k+1}\}$, with $s_1<s_2<\ldots<s_{k+1}$, as support points in the interval $\mathcal{I}_j$. We can build linear functions $\textbf{r}_{j}^{(v)}=[r_{1,j}^{(v)},\ldots,r_{n,j}^{(v)}]$ for every subinterval $\mathcal{I}_{v,v+1}$, using the procedure described in Section \ref{sectBM1}. We recall that this procedure is graphically depicted in Fig. \ref{figcase1}, where we simply need to 
\begin{itemize}
	\item substitute $\mathcal{I}_j$ by $\mathcal{I}_{v,v+1}$ and
	\item when the simple estimate $x_{i,j}\notin \mathcal{I}_{v,v+1}$, substitute $x_{i,j}$ by $s_v$ ($x_{i,j}$ by $s_{v+1}$) if $x_{i,j}<s_v$ (if $x_{i,j}>s_{v+1}$, respectively). 
\end{itemize}

Using $\textbf{r}_{j}^{(v)}$ we compute a bound $\gamma_{j}^{(v)}$, $v=1,\ldots,k$, and then select $\gamma_{j,k}=\min\limits_{v\in\{1,\ldots,k\}}\gamma_{j}^{(v)}$. Note that the subscript $k$ in $\gamma_{j,k}$ indicates how many support points have been used to computed the bound in $\mathcal{I}_j$ (which becomes tighter as $k$ increases). Moreover if we take a new (arbitrary) support point $s^{*}$ from the subinterval ${\mathcal{I}}_{v^*,v^*+1}$ that contains $\gamma_{j,k}$, and extend the set of support points with it, $\mathcal{S}_{j,k+1}=\{s_1,\ldots,s^{*},\ldots, s_{k+2}\}$ with $s_1<s_2<\ldots<s^{*}<\ldots<s_{k+2}$, then we can iterate the proposed procedure and obtain a refined version of the bound, denoted $\gamma_{j,k+1}$. 
   
The proposed iterative algorithm is described, with detail, in Table \ref{ImproveBound}. Note that $k$ is an iteration index that makes explicit the number of support points $s_v$. If we plug this iterative procedure for the computation of $\gamma_{j}$ into BM1 (specifically, replacing steps 3 and 4 of Table \ref{tabla1}), we obtain a new technique that we will hereafter term bounding method 2 (BM2).

\begin{figure*}[htb]
\centering
\centerline{
 		\includegraphics[width=4.1cm,height=3.2cm]{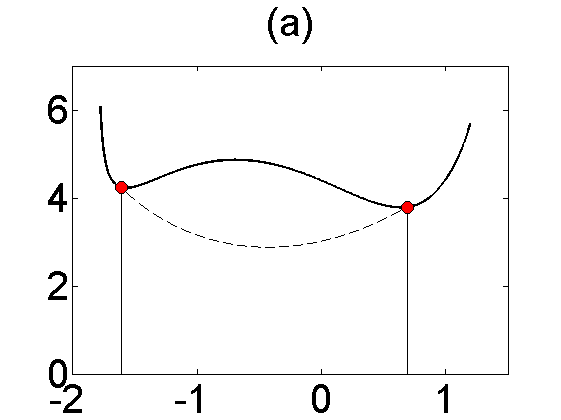}
 		\includegraphics[width=4.1cm,height=3.2cm]{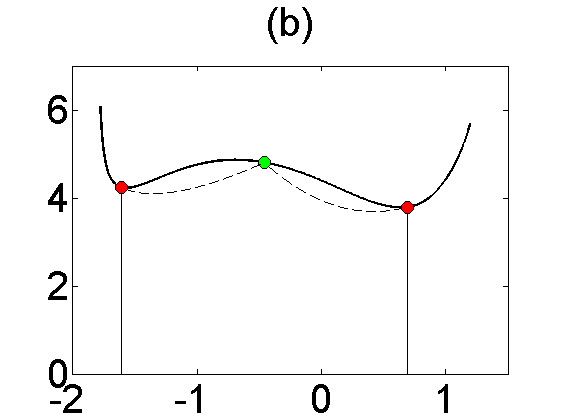}
 		\includegraphics[width=4.1cm,height=3.2cm]{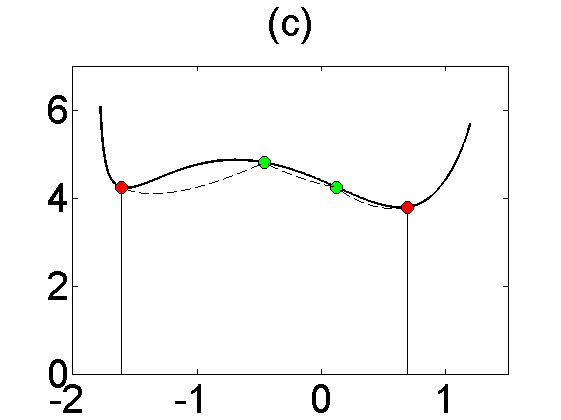}
 		\includegraphics[width=4.1cm,height=3.2cm]{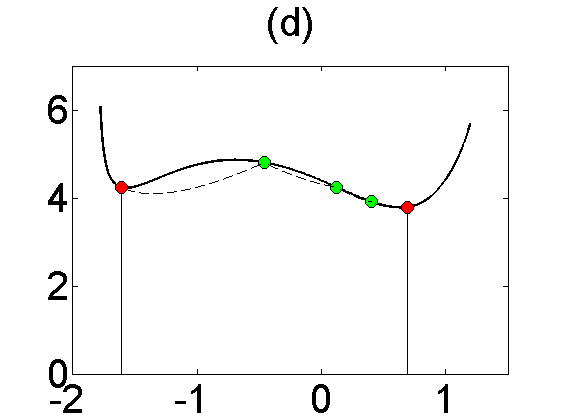}		
 		}
\caption{Four steps of the iterative algorithm choosing $s_{*}$ as the middle point of the subinterval $\mathcal{I}_{v^{*},v^{*}+1}$. The solid line shows the system potential $V(x;\textbf{y},\textbf{g})=(y_{1}-\exp{(x)})^2-\log(y_{2}-\exp{(-x)}+1)+(y_{2}-\exp{(-x)})+1$ (see the example in Section \ref{Example1}), with $y_1=5$ and $y_2=2$, while the dashed line shows the modified potential $V(x;\textbf{y},\textbf{r}_j)$. We start in plot (a) with two points $\mathcal{S}_{j,1}=\{\min{(\mathcal{X}_{j})}, \ \max{(\mathcal{X}_{j})}\}$. At each iteration, we add a new point chosen in the subinterval $\mathcal{I}_{v^{*},v^{*}+1}$ that contains the latest bound. It is apparent that $V(x;\textbf{y},\textbf{r}_j)$ becomes a better approximation of $V(x;\textbf{y},\textbf{g}_j)$ each time we add a new support point.}
\label{ImproveCota}
\end{figure*}

As an illustration, Figure \ref{ImproveCota} shows four steps of the iterative algorithm. In Figure \ref{ImproveCota} (a) there are two support points $\mathcal{S}_{j,1}=\{\min{(\mathcal{X}_{j})}, \ \max{(\mathcal{X}_{j})}\}$, which yield a single interval $\mathcal{I}_{1,2}=\mathcal{I}_{j}$. In Figures \ref{ImproveCota} (b)-(c)-(d), we successively add a point $s^{*}$ chosen in the interval $\hat{\mathcal{I}}_{v^*,v^*+1}$ that contains the latest bound. In this example, the point $s^{*}$ is chosen deterministically as the mean of the extremes of the interval $\mathcal{I}_{v^*,v^*+1}$.  
\begin{table}[!hbt]
\caption{Iterative algorithm to improve $\gamma_{j}$.}
\label{ImproveBound}
\begin{center}
\begin{tabular}{||l||}
\hline
\hline
1. Start with $\mathcal{I}_{1,2}=\mathcal{I}_{j}$, and $\mathcal{S}_{j,1}=\{\min{(\mathcal{X}_{j})}, \ \max{(\mathcal{X}_{j})}\}$. Let $v^{*}=1$ and $k=1$. \\
2.	 Choose an arbitrary interior point $s^{*}$ in $\mathcal{I}_{v^*,v^*+1}$, and update the set of support points $\mathcal{S}_{j,k}=\mathcal{S}_{j,k-1} \cup \ \{s^{*}\}$. \\
3.   Sort $\mathcal{S}_{j,k}$ in ascending order, so that $\mathcal{S}_{j,k}=\{s_{1},\ldots,s_{k+1}\}$ where $s_{1}=\min(\mathcal{X}_{j})$, $s_{k+1}=(\max\mathcal{X}_{j})$, \\ \ \ \ and $k+1$ is the number of elements of $\mathcal{S}_{j,k}$. \\
4.   Build $\textbf{r}^{(v)}_{j}(x)$ for each interval $\mathcal{I}_{v,v+1}=[s_{v},\ s_{v+1}]$ with $v=1,\ldots,k$. \\
5.   Find $\gamma_{j}^{(v)}=\min{V(x;\textbf{y},\textbf{r}_{j}^{(v)})}$, for $v=1,\ldots,k$.\\
6. Set the refined bound $\gamma_{j,k}=\min\limits_{v\in\{1,\ldots,k\}}\gamma_{j}^{(v)}$, and set $v_{*}=\arg\min\limits_{v}\gamma_{j}^{(v)}$.\\
7. To iterate, go back to step 2. \\
\hline
\hline
\end{tabular}
\end{center}
\end{table} 

\subsection{Lower bound $\gamma_{2}$ for quadratic potentials}
\label{sectgamma}
Assume that the joint potential is quadratic, i.e., $V^{(n)}_{2}(y_{1}-g_{1,j}(x),\ldots,y_{n}-g_{n,j}(x))=\sum_{i=1}^{n}(y_{i}-g_{i,j}(x))^{2}$ for each $j=1,\ldots,q$, and construct the set of linearities $r_{i,j}(x)=a_{i,j}x+b_{i,j}$, for $i=1,\ldots,n$ and $j=1,\ldots,q$. The modified system potential in $\mathcal{B}_{j}$ becomes 
\begin{equation}
\label{LS}
V_{2}(x;\textbf{y},\textbf{r}_{j})=\sum_{i=1}^{n}(y_{i}-r_{i,j}(x))^{2}=\sum_{i=1}^{n}(y_{i}-a_{i,j}x-b_{i,j})^{2},
\end{equation}
and it turns out straightforward to compute  $\gamma_{2,j}=\min\limits_{x\in \mathcal{B}_{j}} V(x;\textbf{y},\textbf{r}_{j})$. Indeed, if we denote $\textbf{a}_{j}=[a_{1,j},\ldots,a_{n,j}]^{\top}$ and $\textbf{w}_{j}=[y_{1}-b_{1,j},\ldots,y_{n}-b_{n,j}]^{\top}$, then we can readily obtain  
\begin{eqnarray}
\tilde{x}_{j}=\arg\min_{x\in \mathcal{B}_{j}} V(x;\textbf{y},\textbf{r}_{j}) =\frac{\textbf{a}_{j}^{\top}\textbf{w}_{j}}{\textbf{a}_{j}^{\top}\textbf{a}_{j}},
\end{eqnarray}
and $\gamma_{2,j}=V(\tilde{x}_{j};\textbf{y},\textbf{r}_{j})$. It is apparent that $\gamma_{2}=\min\limits_{j}\gamma_{2,j}\leq V(x;\textbf{y},\textbf{g})$. Furthermore, $\tilde{x}_{j}$ is an approximation of the ML estimator $\hat{x}_j$ restricted to $\mathcal{B}_j$.

\subsection{Adaptation of $\gamma_{2}$ for generic system potentials}
\label{transR}
If the joint potential is not quadratic, in general it can still be difficult to minimize the modified function $V(x;\textbf{y},\textbf{r})$, despite the replacement of the nonlinearities $g_{i,j}(x)$ with the linear functions $r_{i,j}(x)$. In this section, we propose a method to transform the bound for a quadratic potential, $\gamma_2$, into a bound for some other, non-quadratic, potential function.
 
Consider an arbitrary joint potential $V^{(n)}$ and assume the availability of an invertible increasing function $R$ such that $R\circ V^{(n)} \geq V^{(n)}_{2}$, where $\circ$ denotes the composition of functions. Then, for the system potential we can write
\begin{gather}
\begin{split}
\label{quellagiusta}
 (R\circ V)(x;\textbf{y},\textbf{g}) &\geq V_{2}^{(n)}(y_{1}-g_{1}(x),\ldots,y_{n}-g_{n}(x))\\ 
 &= \sum_{i=1}^{n} (y_{i}-g_{i}(x))^{2} \geq \gamma_{2}.
\end{split}
\end{gather}
and, as consequence, $V(x;\textbf{y},\textbf{g}) \geq R^{-1}\left(\gamma_{2}\right)=\gamma$, hence $\gamma$ is a lower bound for the non-quadratic system potential $V(x;\textbf{y},\textbf{g})$ constructed from $V^{(n)}$. 

For instance, consider the family of joint potentials $V^{(n)}_{p}$. Using the monotonicity of $\mathcal{L}^p$ norms, it is possible to prove \cite{Williams91} that 
\begin{equation}
\label{follia1}
\small
\left(\sum_{i=1}^{n} \left|\vartheta_{i}\right|^{p}\right)^{\frac{1}{p}}\geq \left(\sum_{i=1}^{n} \vartheta_{i}^{2}\right)^{\frac{1}{2}}, \ \mbox{for}  \ \ 0\leq p\leq2, \ \mbox{and}
\end{equation}
\begin{equation}
\label{follia2}
\small
n^{\left(\frac{p-2}{2p}\right)}\left(\sum_{i=1}^{n} \left|\vartheta_{i}\right|^{p}\right)^{\frac{1}{p}}\geq \left(\sum_{i=1}^{n} \vartheta_{i}^{2}\right)^{\frac{1}{2}}, \ \mbox{for} \ \ 2\leq p\leq +\infty.
\end{equation}
Let $R_{1}(v)=v^{2/p}$. Since this function is, indeed, strictly increasing, we can transform the inequality (\ref{follia1}) into 
\begin{equation}
	R_1\left(\sum_{i=1}^{n}|y_i-g_i(x)|^{p}\right)\geq \sum_{i=1}^{n}(y_i-g_i(x))^{2},
\end{equation}
which yields
\begin{eqnarray}
 \label{prima} 
\sum_{i=1}^{n}|y_i-g_i(x)|^{p} \geq R_1^{-1}\left(\sum_{i=1}^{n}(y_i-g_i(x))^{2}\right) 
= \left(\sum_{i=1}^{n}(y_i-g_i(x))^{2}\right)^{p/2} \geq \gamma_{2}^{p/2},                      	
\end{eqnarray}
hence the transformation $\gamma_{2}^{p/2}$ of the quadratic bound $\gamma_{2}$ is a lower bound for $V_p^{(n)}$ with $0<p\leq 2$. Similarly, if we let 
$R_{2}(v)=\left(n^{\left(\frac{p-2}{2p}\right)} v^{1/p}\right)^{2}$, the inequality (\ref{follia2}) yields 
\begin{equation}
\label{seconda}
\sum_{i=1}^{n}|y_i-g_i(x)|^{p} \geq R_2^{-1}\left(\sum_{i=1}^{n}(y_i-g_i(x))^{2}\right) 
=\left[n^{\left(-\frac{p-2}{2p}\right)}\left(\sum_{i=1}^{n}(y_i-g_i(x))^{2}\right)^{1/2}\right]^{p} \geq n^{\left(-\frac{p-2}{2}\right)} \gamma_{2}^{p/2},                
\end{equation}
hence the transformation $R_2^{-1}(\gamma_2)=n^{-(p-2)/2} \gamma_{2}^{p/2}$ is a lower bound for $V_p^{(n)}$ when $2\leq p<+\infty$. 

It is possible to devise a systematic procedure to find a suitable function $R$ given an arbitrary joint potential
 $V^{(n)}(\bftheta)$, where $\bftheta\dfn[\vartheta_{1},\ldots,\vartheta_{n}]^T$. Let us define the manifold  $\Gamma_v \dfn \left\{\bftheta\in \mathbb{R}^n: \ \ V^{(n)}(\bftheta)=v\right\}$. We can construct $R$ by assigning $R(v)$ with the maximum of the quadratic potential $\sum_i^n \vartheta_i^2$ when $\bftheta\in \Gamma_v$, i.e., we define 
\begin{equation}
\label{follia3}
\small
	R(v)\dfn\max_{\bftheta\in \Gamma_v} \sum_{i=1}^{n} \vartheta_{i}^2.  
\end{equation} 
Note that (\ref{follia3}) is a constrained optimization problem that can be solved using, e.g., Lagrangian multipliers.

From the definition in (\ref{follia3}) we obtain that, $\forall \textbf{\bftheta}\in \Gamma_v$, $R(v)\geq\sum_{i=1}^{n} \vartheta_{i}^2$. In particular, since $V^{(n)}(\bftheta)=v$ from the definition of $\Gamma_v$, we obtain the desired relationship,   
\begin{equation}
\small
	R\left(V^{(n)}(\vartheta_1,\ldots,\vartheta_n)\right)\geq\sum_{i=1}^{n} \vartheta_{i}^2.
\end{equation}
We additionally need to check whether $R$ is a strictly increasing function of $v$. The two functions in the earlier examples of this Section, $R_1$ and $R_2$, can be readily found using this method. 

\subsection{Convex marginal potentials $\bar{V}_{i}$}
\label{sectConMargPot}
Assume that $\mathcal{A}=\left\{\mathcal{B}_{j}\right\}_{j=1}^{q}$ and that we have already found $r_{i,j}(x)=a_{i,j}x+b_{i,j}$, $i=1,\ldots,n$ and  $j=1,\ldots,q$, using the technique in Section \ref{sectBM1}. If a marginal potential $\bar{V}_{i}(\vartheta_{i})$ is convex, the function $\bar{V}_{i}(y_{i}-r_{i,j}(x))$ is also convex in $\mathcal{B}_{j}$. Indeed, for all $x\in \mathcal{B}_{j}$   
\begin{equation}
\label{compomargpot}
	\frac{d^{2}\bar{V}_{i}(y_{i}-r_{i,j}(x))}{d x^{2}}=\frac{d^{2}r_{i,j}}{d x^{2}}\frac{d\bar{V}_{i}}{d \vartheta_{i}}+\left(\frac{dr_{i,j}}{d x}\right)^{2}\frac{d^{2}\bar{V}_{i}}{d \vartheta_{i}^{2}}=0+a_{i}^{2}\frac{d^{2}\bar{V}_{i}}{d \vartheta_{i}^{2}}\geq 0
\end{equation}
where we have used that $\frac{d^{2}r_{i,j}}{d x^{2}}=0$ (since $r_{i,j}$ is linear).   

As a consequence, if all marginal potentials $\bar{V}_{i}(\vartheta_{i})$ are convex, then the modified system potential, $V(x;\textbf{y},\textbf{r}_{j})=c_{n}+\sum_{i=1}^{n} \bar{V}_{i}(y_{i}-r_{i,j}(x))$, is also convex in $\mathcal{B}_{j}$. This is easily shown using (\ref{compomargpot}), to obtain  
\begin{equation}
\frac{d^{2}V(x;\textbf{y},\textbf{r}_j)}{d x^{2}}= \sum_{i=1}^{n}a_{i}^{2}\frac{d^{2}\bar{V}_{i}}{d \vartheta_{i}^{2}}\geq 0, \ \ \forall x\in \mathcal{B}_{j}.
\end{equation}
Therefore, we can use the tangents to $V(x;\textbf{y},\textbf{r}_{j})$ at the limit points of $\mathcal{I}_{j}$ (i.e, $\min(\mathcal{X}_{j})$ and $\max(\mathcal{X}_{j})$) to find a lower bound for the system potential $V(x;\textbf{y},\textbf{g}_j)$. Figure \ref{LBGeneral} (left) depicts a system potential $V(x;\textbf{y},\textbf{g}_j)$ (solid line), the corresponding modified potential $V(x;\textbf{y},\textbf{r}_j)$ (dotted line) and the two tangent lines at $\min(\mathcal{X}_{j})$ and $\max(\mathcal{X}_{j})$. It is apparent that the intersection of the two tangents yields a lower bound in $\mathcal{B}_{j}$. Specifically, if we let $W(x)$ be the piecewise-linear function composed of the two tangents, then the inequality $V(x;\textbf{y},\textbf{g}_j)\geq V(x;\textbf{y},\textbf{r}_j)\geq W(x)$ is satisfied for all $x\in\mathcal{I}_{j}$.   

\section{Adaptive Rejection Sampling}
\label{secARS}
The adaptive rejection sampling (ARS) \cite{Gilks92} algorithm enables the construction of a sequence of proposal densities, $\left\{\pi_t(x)\right\}_{t\in \mathbb{N}}$, and bounds tailored to the target density. Its most appealing feature is that each time we draw a sample from a proposal $\pi_t$ and it is rejected, we can use this sample to build an improved proposal, $\pi_{t+1}$, with a higher mean acceptance rate. 

Unfortunately, this attractive ARS method can only be applied with target pdf's which are log-concave (hence, unimodal), which is a very stringent constraint for may practical applications. Next, we briefly review the ARS algorithm and then proceed to introduce its extension for non-log-concave and multimodal target densities.

Let $p(x|\textbf{y})$ denote the target pdf\footnote{The method does not require that the target density be a posterior pdf, but we prefer to keep the same notation as in the previous section for coherence.}. The ARS procedure can be applied when $\log[p(x|\textbf{y})]$ is concave, i.e., when the potential function $V(x;\textbf{y},\textbf{g})\dfn-\log[p(x|\textbf{y})]$ is strictly convex. Let $\mathcal{S}_t=\{s_1, s_2,\ldots, s_{k_{t}}\}$ be a set of support points in the domain $D$ of $V(x;\textbf{y},\textbf{g})$. From $\mathcal{S}_t$ we build a piecewise-linear lower hull of $V(x;\textbf{y},\textbf{g})$, denoted $W_t(x)$, formed from segments of linear functions tangent to $V(x;\textbf{y},\textbf{g})$ at the support points in $\mathcal{S}_t$. Figure \ref{LBGeneral} (center) illustrates the construction of $W_t(x)$ with three support points for a generic log-concave potential function $V(x;\textbf{y},\textbf{g})$. 

Once $W_t(x)$ is built, we can use it to obtain an exponential-family proposal density     
\begin{equation}
\label{ARSproposal}
	\pi_{t}(x)=c_{t}\exp[-W_{t}(x)],
\end{equation}  
where $c_t$ is the proportionality constant. Therefore $\pi_t(x)$ is  piecewise-exponential and very easy to sample from. Since $W_t(x)\leq V(x;\textbf{y},\textbf{g})$, we trivially obtain that $\frac{1}{c_t}\pi(x)\geq p(x|\textbf{y})$
and we can apply the RS principle.

When a sample $x'$ from $\pi_t(x)$ is rejected we can incorporate it into the set of support points, $\mathcal{S}_{t+1}=S_t \cup \{x'\}$ (and $k_{t+1}=k_t+1$). Then we compute a refined lower hull, $W_{t+1}(x)$, and a new proposal density $\pi_{t+1}(x)=c_{t+1}\exp\{-W_{t+1}(x)\}$. Table \ref{ARS} summarizes the ARS algorithm.
\begin{table}[!hbt]
\begin{center}
\caption{Adaptive Rejection Sampling Algorithm.}
\label{ARS}
\begin{tabular}{||l||}
\hline
\hline
1. Start with $t=0$, $\mathcal{S}_{0}=\{s_{1}, \ s_{2}\}$ where $s_{1}<s_{2}$, and the derivatives of $V(x,\textbf{y,\textbf{g}})$ in $s_{1},s_{2}\in D$ having different signs. \\
2. Build the piecewise-linear function $W_{t}(x)$ as shown in Figure \ref{LBGeneral} (center), using the tangent lines to $V(x;\textbf{y},\textbf{g})$ \\ \ \ \ at the support points $\mathcal{S}_{t}$.\\
3. Sample $x'$ from $\pi_{t}(x)\propto \exp\{-W_{t}(x)\}$, and $u'$ from $\mathcal{U}([0,1])$. \\
4. If $u'\leq \frac{p(x'|\textbf{y})}{\exp[-W_{t}(x')]}$ accept $x'$ and set $\mathcal{S}_{t+1}=\mathcal{S}_{t}$, $k_{t+1}=k_t$. \\
5. Otherwise, if $u'> \frac{p(x'|\textbf{y})}{\exp[-W_{t}(x')]}$, reject $x'$, set $\mathcal{S}_{t+1}=\mathcal{S}_{t}\cup \{x'\}$ and update $k_{t+1}=k_t+1$. \\ 
6. Sort $\mathcal{S}_{t+1}$ in ascending order, increment $t$ and go back to step 2.\\
\hline
\hline
\end{tabular}
\end{center}
\end{table}

\section{Generalization of the ARS Method}
\label{secMARS}

In this section we introduce a generalization of the standard ARS scheme that can cope with a broader class of target pdf's, including many multimodal distributions. The standard algorithm of \cite{Gilks92}, described in Table \ref{ARS}, is a special case of the method described below.

\subsection{Generalized adaptive rejection sampling}
We wish to draw samples from the posterior $p(x|\textbf{y})$. For this purpose, we assume that 
\begin{itemize}
	\item all marginal potential functions, $\bar{V}_i(\vartheta_i)$, $i=1,\ldots,n$, are strictly convex,
	\item the prior pdf has the form $p(x)\propto \exp\{-\bar{V}_{n+1}(\mu-x)\}$, where $\bar{V}_{n+1}$ is also a convex marginal potential with its mode located at $\mu$, and
	\item the nonlinearities $g_{i}(x)$ are either convex or concave, not necessarily monotonic.
\end{itemize}
We incorporate the information of the prior by defining an extended observation vector,
$\tilde{\textbf{y}}\dfn[y_{1},\ldots,y_{n},y_{n+1}=\mu]^{\top}$, and an extended vector of nonlinearities,  $\tilde{\textbf{g}}(x)\dfn[g_{1}(x),\ldots,g_{n}(x),g_{n+1}(x)=x]^{\top}$. As a result, we introduce the extended system potential function 
\begin{equation}
V(x;\tilde{\textbf{y}},\tilde{\textbf{g}})\dfn V(x;\textbf{y},\textbf{g})+\bar{V}_{n+1}(\mu-x)= -\log[p(x|\textbf{y})]+c_{0},
\end{equation}
where $c_{0}$ accounts for the superposition of constant terms that do not depend on $x$. We remark that the function $V(x;\tilde{\textbf{y}},\tilde{\textbf{g}})$ constructed in this way is not necessarily convex. It can present several minima and, as a consequence, $p(x|\textbf{y})$ can present several maxima.

Our technique is adaptive, i.e., it is aimed at the construction of a sequence of proposals, denoted $\pi_t(x)$, $t\in \mathbb{N}$, but relies on the same basic arguments already exploited to devise the BM1. To be specific, at the $t$-th iteration of the algorithm we seek to replace the nonlinearities $\{g_i\}_{i=1}^{n+1}$ by  piecewise-linear functions $\{r_{i,t}\}_{i=1}^{n+1}$ in such a way that the inequalities 
\begin{equation}
\label{condR12}
\left|y_{i}-r_{i,t}(x)\right| \leq \left|y_{i}-g_{i}(x)\right| \ \ \mbox{and}  	
\end{equation}
\begin{equation}
\label{condR22}
(y_{i}-r_{i,t}(x))(y_{i}-g_{i}(x))\geq 0 
\end{equation}
are satisfied $\forall x\in \mathbb{R}$. Therefore, we repeat the same conditions as in Eqs. (\ref{condR1})-(\ref{condR2}) but the derivation of the generalized ARS (GARS) algorithm does not require the partition of the SoI space, as it was needed for the BM1. 

We will show that it is possible to construct adequate piecewise-linear functions of the form 
\begin{equation}
\label{defRI}
r_{i,t}(x)\dfn	\left\{
\begin{array}{l}
\max[\bar{r}_{i,1}(x),\ldots,\bar{r}_{i,K_t}(x)], \ \ \mbox{if} \ g_i \ \mbox{is convex} \\
\min[\bar{r}_{i,1}(x),\ldots,\bar{r}_{i,K_t}(x)], \ \ \mbox{if} \ g_i \ \mbox{is concave}\\
\end{array}
\right.
\end{equation} 
where $i=1,\ldots,n$ and each $\bar{r}_{i,j}(x)$, $j=1,\ldots,K_{t}$, is a purely linear function. The number of linear functions involved in the construction of $r_{i,t}(x)$ at the $t$-th iteration of the algorithm, denoted $K_t$, determines how tightly $\pi_t(x)$ approximates the true density $p(x|\textbf{y})$ and, therefore, the higher $K_t$, the higher expected acceptance rate of the sampler. In Section \ref{GARSalgo} below, we explicitly describe how to choose the linearities $\bar{r}_{i,j}(x)$, $j=1,\ldots,K_{t}$, in order to ensure that (\ref{condR12}) and (\ref{condR22}) hold. We will also show that, when a proposed sample $x'$ is rejected, $K_t$ can be increased ($K_{t+1}=K_{t}+1$) to improve the acceptance rate. 

Let $\tilde{\textbf{r}}_t\dfn[r_{1,t}(x),\ldots,r_{n,t}(x),r_{n+1,t}(x)=x]^{\top}$ be the extended vector of piecewise-linear functions, that yields the modified potential $V(x;\tilde{\textbf{y}},\tilde{\textbf{r}}_t)$.
The same argument used in Section \ref{sectBM1} to derive the BM1 shows that, if (\ref{condR12}) and (\ref{condR22}) hold, then $V(x;\tilde{\textbf{y}},\tilde{\textbf{r}}_t)\leq V(x;\tilde{\textbf{y}},\tilde{\textbf{g}})$, $\forall x\in \mathbb{R}$. Finally, we build a piecewise-linear lower hull $W_t(x)$ for the modified potential, as explained below, to obtain $W_t(x)\leq V(x;\tilde{\textbf{y}},\tilde{\textbf{r}}_t)\leq V(x;\tilde{\textbf{y}},\tilde{\textbf{g}})$.

The definition of the piecewise-linear function $r_{i,t}(x)$ in (\ref{defRI}) can be rewritten in another form 
\begin{equation}
\label{defR12}
	r_{i,t}(x)\dfn \bar{r}_{i,j}(x) \ \ \mbox{for} \ \ x\in[a,b]
\end{equation}
where $a$ is the abscissa of the intersection between the linear functions $\bar{r}_{i,j-1}(x)$ and $\bar{r}_{i,j}(x)$, and $b$ is the abscissa of the intersection between $\bar{r}_{i,j}(x)$ and $\bar{r}_{i,j+1}(x)$. Therefore, we can define the set of all abscissas of intersection points 
\begin{equation}
	\mathcal{E}_{t}=\{u\in \mathbb{R}:\ \bar{r}_{i,j}(u)=\bar{r}_{i,j+1}(u) \ \ \mbox{for} \ i=1,\ldots,n+1, \ j=1,\ldots,K_t-1 \},
\end{equation}
and sort them in ascending order
\begin{equation}
	u_1<u_2<\ldots<u_Q
\end{equation}
where $Q$ is the total number of intersections. Then 
\begin{itemize}
	\item[a)] since we have assumed that the marginal potentials are convex, we can use Eq. (\ref{defR12}) and the argument of Section \ref{sectConMargPot} to show that the modified function $V(x;\tilde{\textbf{y}},\tilde{\textbf{r}}_t)$ is convex in each interval $[u_{q},u_{q+1}]$, with $q=1,\ldots,Q$, and,    
 \item[b)] as a consequence, we can to build $W_t(x)$ by taking the linear functions tangent to $V(x;\tilde{\textbf{y}},\tilde{\textbf{r}}_t)$ at every intersection point $u_q$, $q=1,\ldots,Q$. 
\end{itemize}

\begin{figure*}[hbt]
\centering
\centerline{
	 \includegraphics[width=4.3cm,height=3.1cm]{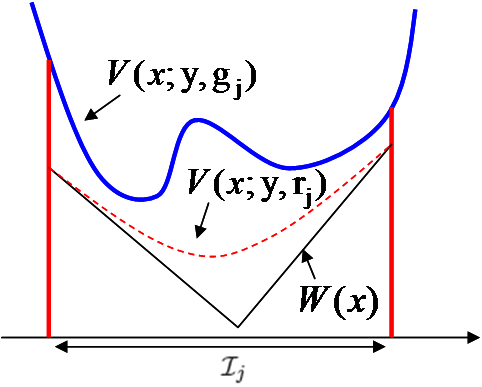}
    \includegraphics[width=4.4cm,height=3.3cm]{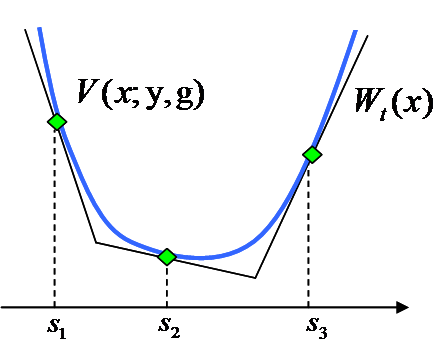}
 		\includegraphics[width=6.3cm,height=3.6cm]{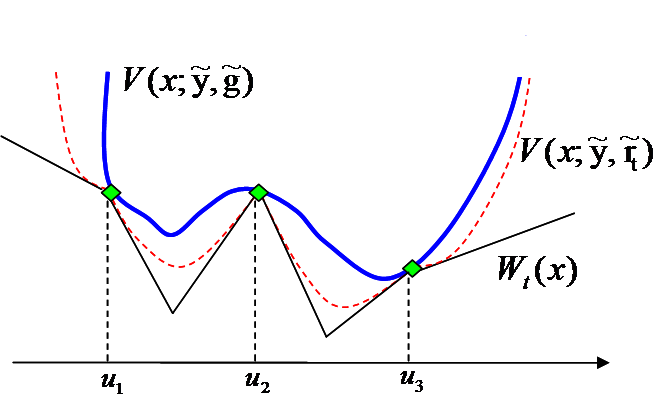}
 		}
\caption{Left: The intersection of the tangents to $V(x;\textbf{y},\textbf{r}_{j})$ (dashed line) at $\min(\mathcal{X}_{j})$ and $\max(\mathcal{X}_{j})$ is a lower bound for $V(x;\textbf{y},\textbf{g}_{j})$ (solid line). Moreover, note that the resulting piecewise-linear function $W(x)$ satisfies the inequality $V(x;\textbf{y},\textbf{g}_j)\geq V(x;\textbf{y},\textbf{r}_j)\geq W(x)$, for all $x\in \mathcal{I}_{j}$. Center: Example of construction of the piecewise-linear function $W_{t}(x)$ with 3 support points $\mathcal{S}_t=\{s_1,s_2,s_3\}$, as carried out in the ARS technique. The function $W_{t}(x)$ is formed from segments of linear functions tangent to $V(x;\textbf{y},\textbf{g})$ at the support points in $\mathcal{S}_t$. Right: Construction of the piecewise linear function $W_{t}(x)$ as tangent lines to the modified potential $V(x;\tilde{\textbf{y}},\tilde{\textbf{r}}_t)$ at three intersections points $u_1$, $u_2$ and $u_3$,  as carried out in the ARS technique.}
\label{LBGeneral}
\end{figure*}

Fig. \ref{LBGeneral} (right) depicts the relationship among $V(x;\tilde{\textbf{y}},\tilde{\textbf{g}})$, $V(x;\tilde{\textbf{y}},\tilde{\textbf{r}}_t)$ and $W_t(x)$. Since $W_t(x)$ is piecewise linear, the corresponding pdf $\pi_t(x)\propto \exp\{-W_{t}(x)\}$ is piecewise exponential and can be easily used in a rejection sampler (we remark that $W_{t}(x)\leq V(x;\tilde{\textbf{y}},\tilde{\textbf{g}})$, hence $\pi_t(x)\propto \exp\{-W_{t}(x)\}\geq \exp\{-V(x;\tilde{\textbf{y}},\tilde{\textbf{g}})\}\propto p(x|\textbf{y})$).

Next subsection is devoted to the derivation of the linear functions needed to construct $\tilde{\textbf{r}}_t$. Then, we describe how the algorithm is iterated to obtain a sequence of improved proposal densities and provide a pseudo-code. Finally, we describe a limitation of the procedure, that yields improper proposals in a specific scenario. 

\subsection{Construction of linear functions $\bar{r}_{i,j}(x)$}  
\label{GARSalgo}
A basic element in the description of the GARS algorithm in the previous section is the construction of the linear functions $\bar{r}_{i,j}(x)$. This issue is addressed below. For clarity, we consider two cases corresponding to non-monotonic and monotonic nonlinearities, respectively. It is important to remark that the nonlinearities $g_i(x)$, $i=1,\ldots,n$ (remember that $g_{n+1}(x)=x$ is linear), can belong to different cases. 

\subsubsection{Non-monotonic nonlinearities}
\label{NewFirstCase}
Assume $g_i(x)$ is a non-monotonic, either concave or convex, function. We have three possible scenarios depending on the number of simple estimates for $g_i(x)$: (a) there exist two simple estimates, $x_{i,1}<x_{i,2}$, (b) there exists a single estimate, $x_{i,1}=x_{i,2}$, or (c) there is no solution for the equation $y_i=g_i(x)$. 

Let us assume that $x_{i,1}<x_{i,2}$ and denote $\mathcal{J}_i\dfn [x_{i,1},x_{i,2}]$. Let us also introduce a set of support points $\mathcal{S}_{t}\dfn\{s_{1},\ldots,s_{k_t}\}$ that contains at least the simple estimates  and an arbitrary point $s\in \mathcal{J}_{i}$, i.e., $x_{i,1},x_{i,2}\in \mathcal{S}_{t}$. The number of support points, $k_t$, determines the accuracy of the approximation of the nonlinearity $g_i(x)$ that can be achieved with the piecewise-linear function $r_{i,t}(x)$. In Section \ref{secSummary} we show how this number increases as the GARS algorithm iterates. Now, we assume it is given and fixed.   
 
\begin{figure*}[hbt]
\centering
\centerline{
 		\includegraphics[width=4.5cm,height=4cm]{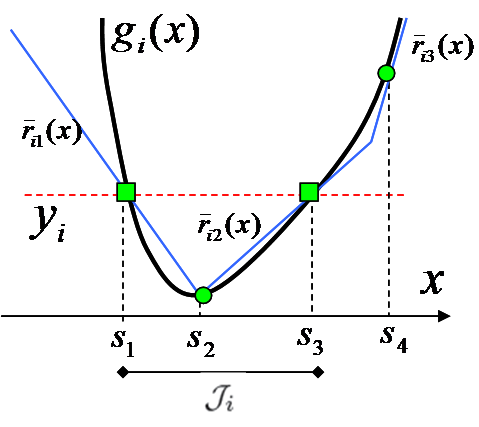}
 		\includegraphics[width=4.5cm,height=3.7cm]{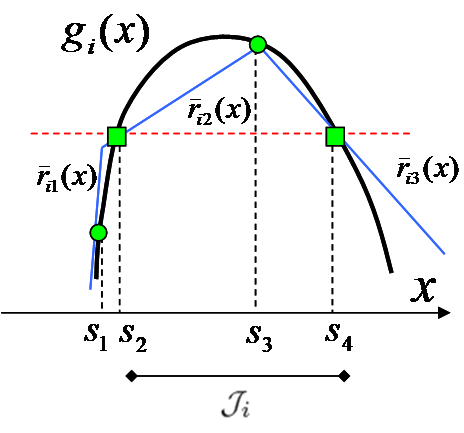}
 		 \includegraphics[width=4.5cm,height=3.5cm]{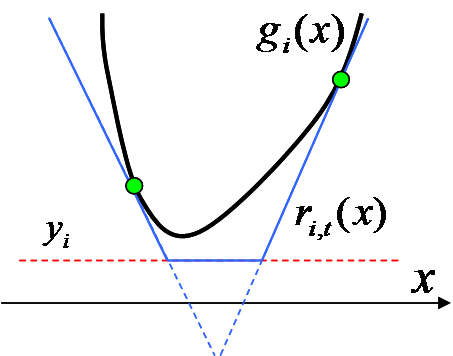}
 		}
\caption{Construction of the piecewise linear function $r_{i,t}(x)$ for non-monotonic functions. The straight lines $\bar{r}_{i,j}(x)$ form a piecewise linear function that is closer to the observation value $y_i$ (dashed line) than the nonlinearity $g_i(x)$, i.e., $\left|y_{i}-r_{i,t}(x)\right| \leq \left|y_{i}-g_{i}(x)\right|$. Moreover, $r_{i,t}(x)$ and $g_i(x)$ are either simultaneously greater than (or equal to) $y_{i}$, or simultaneously lesser than (or equal to) $y_{i}$, i.e., $(y_{i}-r_{i,t}(x))(y_{i}-g_{i}(x))\geq 0$. Therefore, the inequalities (\ref{condR12}) and (\ref{condR22}) are satisfied. The point $(s_j,g_i(s_j))$, corresponding to support point $s_j$, is represented either by a square or a circle, depending on whether it is a simple estimate or not, respectively. Left: construction of $r_{i,t}(x)$ with $k_t=4$ support points when the nonlinearity $g_i(x)$ is convex, therefore $r_{i,t}(x)=\max[\bar{r}_{i,1}(x),\ldots,\bar{r}_{i,3}(x)]$ ($K_t=k_t-1=3$). We use the tangent to $g_{i}(x)$ at $x=s_4$ because $s_4\notin \mathcal{J}_{i}=[s_1,s_3]$, where $s_1=x_{i,1}$ and $s_3=x_{i,2}$ are the simple estimates (represented with squares). Center: since the nonlinearity $g_i(x)$ is concave, $r_{i,t}(x)=\min[\bar{r}_{i,1}(x),\ldots,\bar{r}_{i,3}(x)]$. We use the tangent to $g_{i}(x)$ at $s_4$ because $s_1\notin \mathcal{J}_{i}=[s_2,s_4]$, where $s_2=x_{i,1}$ and $s_4=x_{i,2}$ are the simple estimates (represented with squares). Right: construction of the $r_{i,t}(x)$, with two support points, when there are not simple estimates. We use the tangent lines, but we need a correction in the definition of $r_{i,t}(x)$ in order to satisfy the inequalities (\ref{condR12}) and (\ref{condR22}). Since $g_i(x)$ in the figure is convex, we take $r_{i,t}(x)=\max[\bar{r}_{i,1}(x),\ldots,\bar{r}_{i,K_t}(x),y_i]$.}
\label{BuildLPF}
\end{figure*}

Figure \ref{BuildLPF} illustrates the construction of $\bar{r}_{i,j}(x)$, $j=1,\ldots,K_t$ where $K_t=k_t-1$, and $r_{i,t}(x)$ for a convex nonlinearity $g_i(x)$ (the procedure is completely analogous for concave $g_i(x)$). Assume that the two simple estimates $x_{i,1}<x_{i,2}$ exist, hence $|\mathcal{J}_i|>0$. For each $j\in\{1,\ldots,k_t\}$, the linear function $\bar{r}_{i,j}(x)$ is constructed in one out of two ways: 
\begin{itemize}
	\item[(a)] if $[s_j,s_{j+1}]\subseteq \mathcal{J}_i$, then $\bar{r}_{i,j}(x)$ connects the points $(s_j,g_i(s_j))$ and $(s_{j+1},g_i(s_{j+1}))$, else	 
	\item[(b)] if $s_j\notin \mathcal{J}_i$, then $\bar{r}_{i,j}(x)$ is tangent to $g_{i}(x)$ at $x=s_j$. 
\end{itemize}

From Fig. \ref{BuildLPF} (left and center) it is apparent that $r_{i,t}(x)=\max[\bar{r}_{i,1}(x),\ldots,\bar{r}_{i,K_t}(x)]^{\top}$ built in this way satisfies the inequalities (\ref{condR12}) and (\ref{condR22}), as required. For concave $g_{i}(x)$, (\ref{condR12}) and (\ref{condR22}) are satisfied if we choose $r_{i,t}(x)=\min[\bar{r}_{i,1}(x),\ldots,\bar{r}_{i,K_t}(x)]^{\top}$.  

When $|\mathcal{J}_{i}|=0$ (i.e., $x_{i,1}=x_{i,2}$ or there is no solution for the equation $y_i=g_i(x)$), then each $\bar{r}_{i,j}(x)$ is tangent to $g_i(x)$ at $x=s_j$, $\forall s_j\in \mathcal{S}_t$, and in order to satisfy (\ref{condR12}) and (\ref{condR22}), we need to select  
\begin{equation}
\label{defRIsinpred}
r_{i,t}(x)\dfn	\left\{
\begin{array}{l}
\max[\bar{r}_{i,1}(x),\ldots,\bar{r}_{i,K_t}(x),y_i], \ \ \mbox{if} \ g_i \ \mbox{is convex} \\
\min[\bar{r}_{i,1}(x),\ldots,\bar{r}_{i,K_t}(x),y_i], \ \ \mbox{if} \ g_i \ \mbox{is concave}\\
\end{array}
\right.
\end{equation}

as illustrated in Fig. \ref{BuildLPF} (right).

\subsubsection{Monotonic nonlinearities}
\label{NewSecondCase}
In this case $g_{i}(x)$ is invertible and there are two possibilities: there exists a single estimate, $x_i=g_i^{-1}(y_i)$, or there is no solution for the equation $y_i=g_i(x)$ (where $y_i$ does not belong to the range of $g_i(x)$). Similarly to the construction in Section 
\ref{sectBM1}, we distinguish two cases:
\begin{itemize}
	\item[(a)] if $\frac{d g_{i}(x)}{d x}\times \frac{d^{2}g_{i}(x)}{d x^{2}}\geq 0$, then we define $\mathcal{J}_{i}\dfn(-\infty,x_i]$, and 
	\item[(b)] if $\frac{d g_{i}(x)}{d x}\times \frac{d^{2}g_{i}(x)}{d x^{2}}\leq 0$, then we define $\mathcal{J}_{i}\dfn[x_i,+\infty)$. 
\end{itemize}

The set of support points is $\mathcal{S}_{t}\dfn\{s_{1},\ldots,s_{k_t}\}$, with $s_{1}<s_{2}\ldots<s_{k_t}$, and includes at least the simple estimate $x_i$ and an arbitrary point $s\in \mathcal{J}_{i}$, i.e., $x_{i},s\in \mathcal{S}_{t}$. 

\begin{figure*}[hbt]
\centering
\centerline{
 		\includegraphics[width=5cm,height=4.1cm]{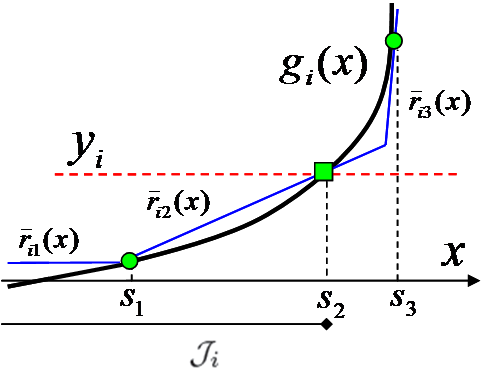}
 		\includegraphics[width=5cm,height=4.6cm]{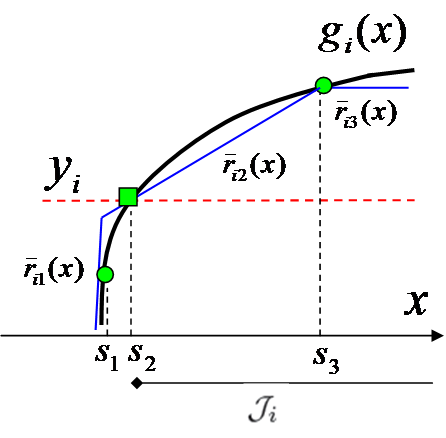}
 		}
\caption{Examples of construction of the piecewise-linear function $r_{i,t}(x)$ with $k_t=3$ support points $s_j$, for the two subcases. It is apparent that $\left|y_{i}-r_{i,t}(x)\right| \leq \left|y_{i}-g_{i}(x)\right|$ and that $r_{i,t}(x)$ and $g_i(x)$ are either simultaneously greater than (or equal to) $y_{i}$, or simultaneously lesser than (or equal to) $y_{i}$, i.e., $(y_{i}-r_{i,t}(x))(y_{i}-g_{i}(x))\geq 0$. The simple estimates are represented by squares while all other support points are drawn as circles. Left: the figure corresponds to the subcase 1 where $r_{i,t}(x)=\max[\bar{r}_{i,1}(x),\ldots,\bar{r}_{i,3}(x)]$ ($K_t=k_t=3$). Right: the figure corresponds to to the subcase 2 where $r_{i,t}(x)=\min[\bar{r}_{i,1}(x),\ldots,\bar{r}_{i,3}(x)]$ ($K_t=k_t=3$).}
\label{BuildLPF2}
\end{figure*}

The procedure to build $\bar{r}_{i,j}(x)$, for $j=1,\ldots,K_t$, with $K_t=k_t$, is similar to Section \ref{NewFirstCase}. Consider case (a) first. For each $j\in \{2,\ldots,k_t\}$, if $[s_{j-1},s_{j}]\subset \mathcal{J}_{i}=(-\infty,x_i]$, then $\bar{r}_{i,j}(x)$ is the linear function that connects the points $(s_{j-1},g_i(s_{j-1}))$ and $(s_{j},g_i(s_{j}))$. Otherwise, if $s_j\notin \mathcal{J}_{i}=(-\infty,x_i]$, $\bar{r}_{i,j}(x)$ is tangent to $g_i(x)$ at $x=s_j$. Finally, we set $\bar{r}_{i,1}(x)=g_i(s_1)$ for all $x\in \mathbb{R}$. The piecewise linear function $r_{i,t}$ is $r_{i,t}(x)=\max[\bar{r}_{i,1}(x),\ldots,\bar{r}_{i,K_t}(x)]$. This construction is depicted in Fig. \ref{BuildLPF2} (left). 

Case (b) is similar. For each $j\in\{1,\ldots,k_t\}$, if $[s_j,s_{j+1}]\subset \mathcal{J}_{i}=[x_i,+\infty)$, then $\bar{r}_{i,j}(x)$ is the linear function that connects the points $(s_j,g_i(s_j))$ and $(s_{j+1},g_i(s_{j+1}))$.
Otherwise, if $s_j\notin \hat{\mathcal{I}}_{i}=[x_i,+\infty)$, $\bar{r}_{i,j}(x)$ is tangent to $g_i(x)$ at $x=s_j$. Finally, we set $\bar{r}_{i,k_t}(x)=g_i(s_{k_t})$ (remember that, in this case, $K_t=k_t$), for all $x\in \mathbb{R}$. The piecewise linear function $r_{i,t}$ will be $r_{i,t}(x)=\min[\bar{r}_{i,1}(x),\ldots,\bar{r}_{i,K_t}(x)]$. This construction is depicted in Fig. \ref{BuildLPF2} (right). 

It is straightforward to check that the inequalities (\ref{condR12}) and (\ref{condR22}) are satisfied. Note that, in this case, the number of linear functions $\bar{r}_{i,j}(x)$ coincides with the number of support points. If there is not solution for the equation $y_i=g_i(x)$ ($y_i$ does not belong to the range of $g_i(x)$), then (\ref{condR12}) and (\ref{condR22}) are satisfied if we use (\ref{defRIsinpred}) to build $r_{i,t}(x)$.  

\subsection{Summary}
\label{secSummary}
We can combine the elements described in Sections \ref{NewFirstCase} and \ref{NewSecondCase} into an adaptive algorithm that improves the proposal density $\pi_t(x)\propto\exp\{-W_t(x)\}$ each time a sample is rejected.
 
Let $\mathcal{S}_t$ denote the set of support points after the $t$-th iteration. We initialize the algorithm with $\mathcal{S}_{0}\dfn\left\{s_{j}\right\}_{j=1}^{k_0}$ such that
\begin{itemize}
	\item all simple estimates are contained in $\mathcal{S}_{0}$, and
	\item for each interval $\mathcal{J}_i$, $i=1,\ldots,n+1$ , with non-zero length ($|\mathcal{J}_i|>0$), there is at least one (arbitrary) support point contained in $\mathcal{J}_i$. 
\end{itemize}
The proposed GARS algorithm is described in Table \ref{MARS}. Note that every time a sample $x'$ drawn from $\pi_t(x)$ is rejected, $x'$ is incorporated as a support point in the new set $\mathcal{S}_{t+1}=\mathcal{S}_{t}\cup \{x'\}$ and, as a consequence, a refined lower hull $W_{t+1}(x)$ is constructed yielding a better approximation of the system potential function. In this way, $\pi_{t+1}(x)\propto \exp\{-W_{t+1}(x)\}$ becomes closer to $p(x|\textbf{y})$ and it can be expected that the acceptance rate be higher. This is specifically shown in the simulation example in Section \ref{secExample2}.

\begin{table}[!hbt]
\caption{Steps of Generalized Adaptive Rejection Sampling.}
\label{MARS}
\begin{center}
\begin{tabular}{||l||}
\hline
\hline
1. Start with $t=0$ set $\mathcal{S}_{0}\dfn\left\{s_{j}\right\}_{j=1}^{k_0}$.  \\
2. Build $\bar{r}_{i,j}(x)$ for $i=1,\ldots,n+1$, $j=1,\ldots,K_t$, where $K_t=k_t-1$ or $K_t=k_t$ depending on whether $g_i(x)$ is \\ \ \ \ non-monotonic or monotonic, respectively. \\
3. Calculate the set of intersection points $\mathcal{E}_t\dfn\{u\in \mathbb{R}:\ \bar{r}_{i,j}(u)=\bar{r}_{i,j+1}(u) \ \ \mbox{for} \ i=1,\ldots,n+1, \ j=1,\ldots,K_t-1\}$. \\ \ \ \ Let $Q=|\mathcal{E}_t|$  be the number of elements in $\mathcal{E}_t$. \\ 
4. Build $W_{t}(x)$ using the tangent lines to $V(x;\tilde{\textbf{y}},\tilde{\textbf{r}}_t)$ at the points $u_{q}\in \mathcal{E}_t$, $q=1,\ldots,Q$. \\
5. Draw a sample $x'$ from $ \pi_{t}(x)\propto \exp[-W_{t}(x)]$. \\
6. Sample $u'$ from $\mathcal{U}([0,1])$. \\
7. If $u'\leq \frac{p(x'|\textbf{y})}{\exp[-W_{t}(x')]}$ accept $x'$ and set $\mathcal{S}_{t+1}=\mathcal{S}_{t}$. \\
8. Otherwise, if $u'> \frac{p(x'|\textbf{y})}{\exp[-W_{t}(x')]}$ reject $x'$ and update $\mathcal{S}_{t+1}=\mathcal{S}_{t}\cup \{x'\}$. \\
9. Sort $\mathcal{S}_{t+1}$ in ascending order, set $t=t+1$ and go back to step 2.\\
\hline
\hline
\end{tabular}
\end{center}
\end{table} 

\subsection{Improper proposals}
The GARS algorithm as described in Table \ref{MARS} breaks down when every $g_{i}(x)$, $i=1,\ldots,n+1$, is nonlinear and convex (or concave) monotonic. In this case, the proposed construction procedure yields a piecewise lower hull $W_{t}(x)$ which is positive and constant in an interval of infinite length. Thus, the resulting proposal, $\pi_t(x)\propto \exp\{-W_t(x)\}$ is improper ($\int_{-\infty}^{+\infty} \pi_t(x)dx\rightarrow +\infty$) and cannot be used for RS. One practical solution is to substitute the constant piece of $W_t(x)$ by a linear function with a  small slope. In that case, $\pi_t(x)$ is proper but we cannot guarantee that the samples drawn using the GARS algorithm come exactly from the target pdf. Under the assumptions in this paper, however, $g_{n+1}(x)=x$ is linear (due to our choice of the prior pdf), and this is enough to guarantee that $\pi_t(x)$ be proper.

\section{Examples}
\label{sectExample}
\subsection{Example 1: Calculation of upper bounds for the likelihood function}
\label{Example1}
Let $X$ be a scalar SoI with prior density $X\sim p(x)=N(x;0,2)$ and the random observations 
\begin{equation}
\label{sistemaejemplo}
Y_{1}=\exp{(X)}+\Theta_{1}, \ \ Y_{2}=\exp{(-X)}+\Theta_{2},
\end{equation}   
where $\Theta_{1}$, $\Theta_{2}$ are independent noise variables. Specifically, $\Theta_{1}$ is Gaussian noise with $N(\vartheta_1;0,1/2) = k_{1}\exp\left\{-(\vartheta_{1})^{2}\right\}$, and $\Theta_{2}$ has a gamma pdf, $\Theta_{2}\sim\Gamma(\vartheta_2;\theta,\lambda)= k_{2}\vartheta_{2}^{\theta-1} \exp\left\{-\lambda\vartheta_{2}\right\}$, with parameters $\theta=2, \lambda=1$.

The marginal potentials are $\bar{V}_{1}(\vartheta_{1})=\vartheta_{1}^{2}$ and $\bar{V}_{2}(\vartheta_{2})=-\log(\vartheta_{2})+\vartheta_{2}$. Since the minimum of $\bar{V}_{2}(\vartheta_{2})$ occurs in $\vartheta_{2}=1$, we replace $Y_2$ with the shifted observation $Y_{2}^{*}=\exp{(-X)}+\Theta_{2}^{*}$, where $Y_{2}^{*}=Y_{2}-1$, $\Theta_{2}^{*}=\Theta_{2}-1$. Hence, the marginal potential becomes $\bar{V}_{2}(\vartheta_{2}^{*})=-\log(\vartheta_{2}^{*}+1)+\vartheta_{2}^{*}+1$, with a minimum at $\vartheta_2^*=0$, the vector of observations is $\textbf{Y}=[Y_{1},Y_{2}^{*}]^{\top}$ and the vector of nonlinearities is $\textbf{g}(x)=[\exp{(x)},\exp{(-x)}]^{\top}$. 
Due to the monotonicity and convexity of $g_{1}$ and $g_2$, we can work with a partition of $\mathbb{R}$ consisting of just one set, $\mathcal{B}_{1}\equiv \mathbb{R}$. The joint potential is $V^{(2)}(\vartheta_{1},\vartheta_{2}^{*})=\sum_{i=1}^{2}\bar{V}_{i}(\vartheta_{i})=\vartheta_{1}^2-\ln(\vartheta_{2}^{*}+1)+\vartheta_{2}^{*}+1$ and the system potential is
\begin{gather}
\begin{split}
V(x;\textbf{y},\textbf{g})&=V^{(2)}(y_{1}-\exp{(x)},y_{2}^{*}-\exp{(-x)})=  \\
&=(y_{1}-\exp{(x)})^2-\log(y_{2}^{*}-\exp{(-x)}+1)+(y_{2}^{*}-\exp{(-x)})+1.	
\end{split}
\end{gather}
Assume that, $\textbf{Y}=\textbf{y}=[2,5]^{\top}$. The simple estimates are $\mathcal{X}=\{x_{1}=\log(2),x_{2}=-\log(5)\}$, and, therefore, we can restrict the search of the bound to the interval $\mathcal{I}=[\min(\mathcal{X})=-\log(5),\max(\mathcal{X})=\log(2)]$ (note that we omit the subscript because we have just one set, $\mathcal{B}_{1}\equiv \mathbb{R}$). Using the BM1 technique in Section \ref{sectBM1}, we find the linear functions $r_{1}(x)=0.78x+1.45$ and $r_{2}(x)=-1.95x+1.85$.
 
In this case, we can analytically minimize the modified system potential, to obtain $\tilde{x}=-0.4171=\arg\min\limits_{x\in \mathcal{I}}V(x,\textbf{y},\textbf{r})$. The associated lower bound is $\gamma=V(\tilde{x},\textbf{y},\textbf{r})=2.89$ (the true global minimum of the system potential is $3.78$). We can also use the technique in Section \ref{transR} with $R^{-1}(v)=-\log(\sqrt{v}+1)+\sqrt{v}+1$. The lower bound for the quadratic potential is $\gamma_{2}=2.79$ and we can readily compute a lower bound $\gamma=R^{-1}(\gamma_{2})=1.68$ for $V(x;\textbf{y},\textbf{g})$. Since the marginal potentials are both convex, we can also use the procedure described in Section \ref{sectConMargPot}, obtaining the lower bound $\gamma=1.61$. 

\begin{figure*}[htb]
\centering
\centerline{
 		\includegraphics[width=5.3cm,height=4cm]{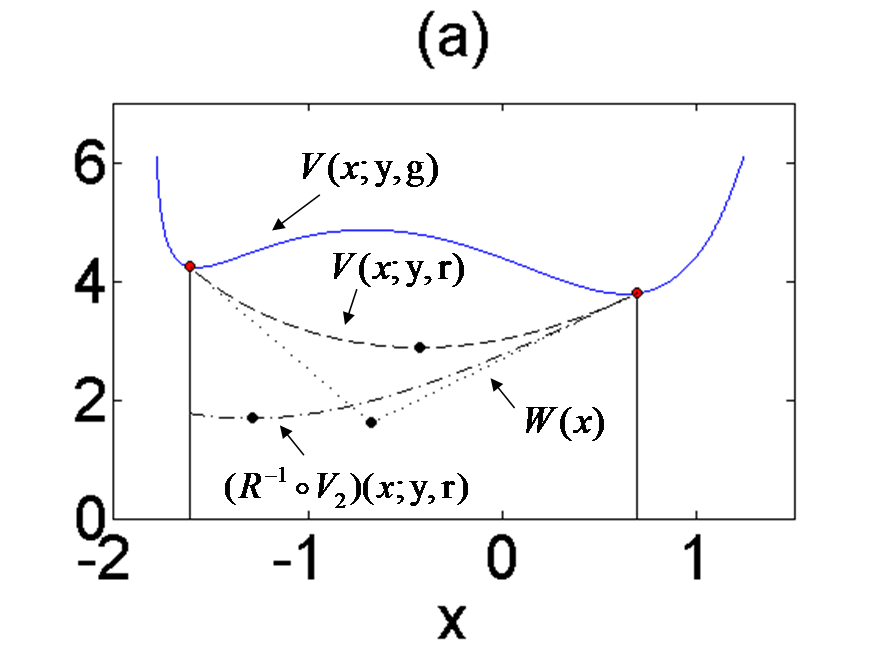}
    \hspace{-0.65cm}	
    \includegraphics[width=5.3cm,height=4cm]{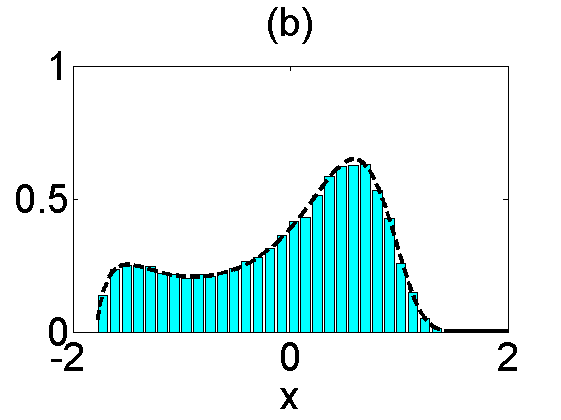}
 		\hspace{-0.65cm}	
 		\includegraphics[width=5.3cm,height=4cm]{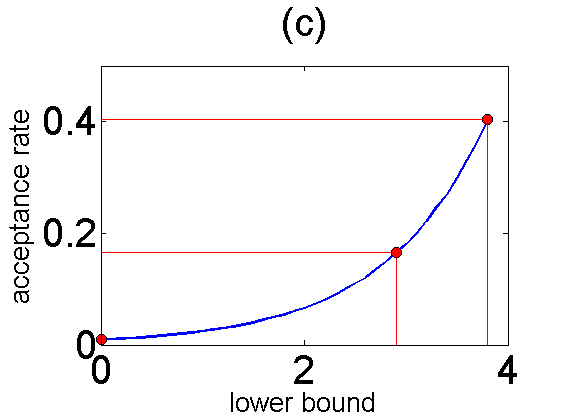}
}
\caption{(a) The system potential $V(x,\textbf{y},\textbf{g})$ (solid), the modified system potential $V(x,\textbf{y},\textbf{r})$ (dashed), function $(R^{-1}\circ V_2)(x,\textbf{y},\textbf{r})$ (dot-dashed) and the  piecewise-linear function $W(x)$ formed by the two tangent lines to $V(x,\textbf{y},\textbf{r})$ at $\min(\mathcal{X})$ and $\max(\mathcal{X})$ (dotted). The corresponding bounds are marked with dark circles. (b) The target density $p(x|\textbf{y})\propto p(\textbf{y}|x)p(x)$ (dashed) and the normalized histogram of $N=10,000$ samples using RS with the the calculated bound $L$. (c) The curve of acceptance rates (averaged over $10,000$ simulations) as a function of the lower bound $\gamma$. The acceptance rate is $1.1\%$ for the trivial bound $\gamma=0$, $18\%$ with $\gamma=2.89$, approximately $40\%$ with $\gamma=3.77$ and $41\%$ with the optimal bound $\gamma=3.78$.} 
	\label{lowerbound}
\end{figure*}

Figure \ref{lowerbound} (a) depicts the system potential $V(x;\textbf{y},\textbf{g})$, and the lower bounds obtained with the three methods. It is the standard BM1 algorithm that yields the best bound.

In order to improve the bound, we can use the iterative BM2 technique described in Section \ref{improve}. With only $3$ iterations of BM2, and minimizing analytically the modified potential $V(x,\textbf{y},\textbf{r})$, we find a very tight lower bound $\gamma=\min\limits_{x\in \mathcal{I}}(V(x,\textbf{y},\textbf{r}))=3.77$ (recall that the optimal bound is $3.78$). Table \ref{tableExample0} summarizes the bounds computed with the different techniques. 

Next, we implement a rejection sampler, using the prior pdf $p(x)=N(x;0,2)\propto \exp\{-x^2/4\}$ as a proposal function and the upper bound for the likelihood $L=\exp\{-3.77\}$. The posterior density has the form  
\begin{equation}
	p(x|\textbf{y})\propto p(\textbf{y}|x)p(x)=\exp\{-V(x;\textbf{y},\textbf{g})-x^2/4\}.
\end{equation}

Figure \ref{lowerbound} (b) shows the normalized histogram of $N=10,000$ samples generated by the RS algorithm, together with the true target pdf $p(x|\textbf{y})$ depicted as a dashed line. The histogram follows closely the shape of the true posterior pdf. Figure \ref{lowerbound} (c)
shows the acceptance rates (averaged over $10,000$ simulations) as a function of the bound $\gamma$. We start with the trivial lower bound $\gamma=0$ and increase it progressively, up to the global minimum $\gamma=3.78$. The resulting acceptance rates are $1.1\%$ for the trivial bound $\gamma=0$, $18\%$ with $\gamma=2.89$ (BM1) and approximately $40\%$ with $\gamma=3.77$ (BM2). Note that the acceptance rate is $\approx 41\%$ for the optimal bound and we cannot improve it any further. This is an intrinsic drawback of a rejection sampler with constant bound $L$ and the principal argument that suggests the use of adaptive procedures.       
 
\begin{table}[!hbt]
\caption{Lower bounds of the system potential function.}
\label{tableExample0}
\vspace{-0.5cm}
\centering
\begin{tabular}{|c|c|c|c|c|c|}
\hline
\hline
Method & BM1 & BM1 + trasformation $R$ & BM1 + tangent lines & BM2 & Optimal Bound\\
\hline
Lower Bound $\gamma$ & 2.89 & 1.68 & 1.61 & 3.77 & 3.78\\       
\hline
\end{tabular}
\end{table}
\vspace{-0.5cm}

\subsection{Example 2: Comparison of ARMS and GARS techniques}
\label{secExample2}
Consider the problem of sampling a scalar random variable $X$ from a posterior bimodal density $p(x|y)\propto p(y|x)p(x)$, where the likelihood function is $p(y|x)\propto\exp\{-\cosh(y-x^{2})\}$ (note that we have a single observation $Y=y_1$) and prior pdf is $p(x)\propto\exp\{-\alpha(\eta-\exp(|x|))^2\}$, with constant parameters $\alpha>0$ and $\eta$. Therefore, the posterior pdf is $p(x|y)\propto\exp\left\{-V(x;\tilde{\textbf{y}},\tilde{\textbf{g}})\right\}$, where $\tilde{\textbf{y}}=[y,\eta]^{\top}$, $\tilde{\textbf{g}}(x)=[g_{1}(x),g_{2}(x)]^{\top}=[x^2,\exp(|x|)]^{\top}$ and the extended system potential function becomes
\begin{equation}
V(x;\tilde{\textbf{y}},\tilde{\textbf{g}})=\cosh(y-x^{2})+\alpha(\eta-\exp(|x|))^2.
\end{equation}
The marginal potentials are $\bar{V}_1(\vartheta_{1})=\cosh(\vartheta_{1})$ and $\bar{V}_2(\vartheta_{2})=\alpha\vartheta_{2}^2$. Note that the density $p(x|y)$ is an even function, $p(x|y)=p(-x|y)$, hence it has a zero mean, $\mu=\int xp(x|y)dx=0$. The constant $\alpha$ is a scale parameter that allows to control the variance of the random variable $X$, both {\it a priori} and {\it a posteriori}. The higher the value of $\alpha$, the more skewed the modes of $p(x|y)$ become. 
   
There are no standard methods to sample directly from $p(x|y)$. Moreover, since the posterior density $p(x|y)$ is bimodal, the system potential is non-log-concave and the ARS technique cannot be applied. However, we can easily use the GARS technique. If, e.g., $\tilde{\textbf{Y}}=\tilde{\textbf{y}}=[y=5,\eta=10]^{\top}$ the simple estimates corresponding to $g_{1}(x)$ are $x_{1,1}=-\sqrt{5}$ and $x_{1,2}=\sqrt{5}$, so that $\mathcal{J}_1=[-\sqrt{5},\sqrt{5}]$. In the same way, the simple estimates corresponding to $g_{2}(x)$ are $x_{2,1}=-\log(10)$ and $x_{2,2}=\log(10)$, therefore $\mathcal{J}_2=[-\log(10),\log(10)]$. 

An alternative possibility to draw from this density is to use the ARMS method \cite{Gilks95}. Therefore, in this section we compare the two algorithms. Specifically, we look into the accuracy in the approximation of the posterior mean $\mu=0$ by way of the sample mean estimate, $\hat{\mu}=\frac{1}{N}\sum_{i=1}^{N}x^{(i)}$, for different values of the scale parameter $\alpha$.  

In particular, we have considered ten equally spacial values of $\alpha$ in the interval $[0.2,5]$ and then performed $10,000$ independent simulations for each value of $\alpha$, each simulation consisting of drawing $5,000$ samples with the GARS method and the ARMS algorithm. Both techniques can be sensitive to their initialization. The ARMS technique starts with $5$ points selected randomly in $[-3.5,3.5]$ (with uniform distribution).  The GARS starts with the set of support points $\mathcal{S}_{0}=\{x_{2,1},x_{1,1},s,x_{1,2},x_{2,2}\}$ sorted in ascending order, including all simple estimates and an arbitrary point $s$ needed to enable the construction in Section \ref{GARSalgo}. Point $s$ is randomly chosen in each simulation, with uniform pdf in $\mathcal{J}_1=[x_{1,1},x_{1,2}]$.  

The simulation results show that the two techniques attain similar performance when $\alpha\in[0.2,1]$ (the modes of $p(x|y)$ are relatively flat). When $\alpha\in[1,4]$ the modes become more skewed and Markov chain generated by the ARMS algorithm remains trapped at one of the two modes in $\approx 10\%$ of the simulations. When $\alpha\in[4,5]$ the same problem occurs in $\approx 25\%$ of the simulations. The performance of the GARS algorithm, on the other hand, is comparatively insensitive to the value of $\alpha$.      

\begin{figure*}[htb]
\centering
\centerline{
    \includegraphics[width=5.3cm,height=4cm]{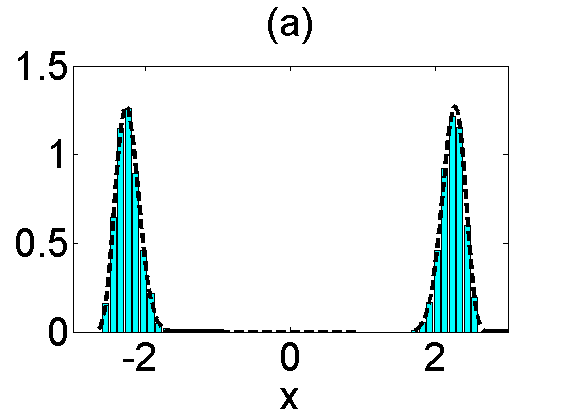}
    \includegraphics[width=5.3cm,height=4cm]{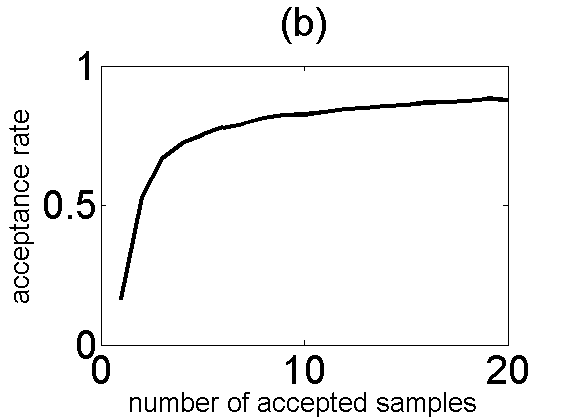}
}
\caption{(a) The bimodal density $p(x|\textbf{y})\propto \exp\left\{-V(x;\tilde{\textbf{y}},\tilde{\textbf{g}})\right\}$ (dashed line) and the normalized histogram of $N=5000$ samples obtained using GARS algorithm. (b) The curve of acceptance rates (averaged over $10,000$ simulations) as a function of the accepted samples.} 
	\label{GARSejemploFig}
\end{figure*}

Figure \ref{GARSejemploFig} (a) shows the posterior density $p(x|y)\propto \exp\left\{-\cosh(y_1-x^{2})-\alpha(\mu-\exp(|x|))^2\right\}$ with $\alpha=0.2$ depicted as a dashed line, and the normalized histogram obtained with the GARS technique. Figure \ref{GARSejemploFig} (b) illustrates the acceptance rates (averaged over $10,000$ simulations) for the first $20$ accepted samples drawn with the GARS algorithm. Every time a sample $x'$ drawn from $\pi_t(x)$ is rejected, it is incorporated as a support point. Then, the proposal pdf $\pi_{t}(x)$ becomes closer to target pdf $p(x|y)$ and, as a consequence, the acceptance rate becomes higher. For instance, the acceptance rate for the first sample is $\approx 16\%$, but for the second sample, it is already $\approx 53\%$. The acceptance rate for the $20$-th sample is $\approx 90\%$. 

\begin{table}[!hbt]
\caption{Estimated posterior mean, $\hat{\mu}$ (for $\alpha=5$).}
\label{tableExample1}
\centering
\begin{tabular}{|c|c|c|c|c|c|}
\hline
\hline
Simulation & 1 & 2 & 3 & 4 & 5 \\
\hline
ARMS & -2.2981 & 0.0267 & 0.0635 & 0.0531 & 2.2994 \\       
\hline
GARS & 0.0772 & -0.0143 & 0.0029 & 0.0319 &  0.0709\\  
\hline
\end{tabular}
\end{table}
\vspace{-0.5cm}

\subsection{Example 3: Target localization with a sensor network}
\label{Example4}
In order to show how the proposed techniques can be used to draw samples from a multivariate (non-scalar) SoI, we consider the problem of positioning a target in a $2$-dimensional space using range measurements. This is a problem that appears frequently in localization applications using sensor networks \cite{Ali07}. 

We use a random vector $\textbf{X}=[X_1,X_2]^{\top}$ to denote the target position in the plane $ \mathbb{R}^{2}$. The prior density of $X$ is $p(x_1,x_2)=p(x_1)p(x_2)$, where $p(x_i)=N(x_i;0,1/2)=k\exp\left\{-(x_i)^{2}\right\}$, $i=1,2$, i.e., the coordinate $X_1$ and $X_2 $ are i.i.d. Gaussian. The range measurements are obtained from two sensor located at $\textbf{h}_1=[0,0]^{\top}$ and $\textbf{h}_2=[2,2]^{\top}$, respectively. The effective observations are the (square) Euclidean distances from the target to the sensors, contaminated with Gaussian noise, i.e.,    
\begin{gather}
\label{sistemaejemplo}
\begin{split}
&Y_{1}=X_1^2+X_2^2+\Theta_{1}, \\
&Y_{2}=(X_1-2)^2+(X_2-2)^2+\Theta_{2},
\end{split}   
\end{gather}   
where $\Theta_{i}$, $i=1,2$, are independent Gaussian variables with identical pdf's, $N(\vartheta_i;0,1/2) = k_{i}\exp\left\{-\vartheta_{i}^{2}\right\}$. Therefore, the marginal potentials are quadratic,  $\bar{V}_i(\vartheta_{i})=\vartheta_{i}^2$, $i=1,2$. The random observation vector is denoted $Y=[Y_1,Y_2]^{\top}$. We note that one needs three range measurements to uniquely determine the position of a target in the plane, so the posterior pdf $p(\textbf{x}|\textbf{y})\propto p(\textbf{y}|\textbf{x})p(\textbf{x})$ is bimodal. 
 
We apply the Gibbs sampler to draw $N$ particles $\textbf{x}^{(i)}=[x_1^{(i)},x_2^{(i)}]^{\top}$, $i=1,\ldots,N$, from the posterior density $p(\textbf{x}|\textbf{y})\propto p(\textbf{y}|x_1,x_2)p(x_1)p(x_2)$. The algorithm can be summarized as follows:
\begin{enumerate}
  \item Set $i=1$, and draw $x_2^{(1)}$ from the prior pdf $p(x_2)$. 
	\item Draw a sample $x_1^{(i)}$ from the conditional pdf $p(x_1|\textbf{y},x_2^{(i)})$, and set $\textbf{x}^{(i)}=[x_1^{(i)},x_2^{(i)}]^{\top}$.
	\item  Draw a sample $x_2^{(i+1)}$ from the conditional pdf $p(x_2|\textbf{y},x_1^{(i)})$.
\item Increment $i=i+1$. If $i>N$ stop, else go back to step 2.
\end{enumerate}
The Markov chain generated by the Gibbs sampler converges to a stationary distribution with pdf $p(x_1,x_2|\textbf{y})$. 

In order to use Gibbs sampling, we have to be able to draw from the conditional densities $p(x_1|\textbf{y},x_2^{(i)})$ and $p(x_2|\textbf{y},x_1^{(i)})$. In general, these two conditional pdf's can be non-log-concave and can have several modes. Specifically, the density $p(x_1|\textbf{y},x_2^{(i)})\propto p(\textbf{y}|x_1,x_2^{(i)})p(x_1)$ can be expressed as $p(x_1|\textbf{y},x_2^{(i)})\propto \exp\{-V(x_1;\tilde{\textbf{y}}_1,\tilde{\textbf{g}}_1)\}$ where $\tilde{\textbf{y}}_1=[y_1-(x_2^{(i)})^2,y_2-(x_2^{(i)}-2)^2,0]^{\top}$, $\tilde{\textbf{g}}_1(x)=[x^2,(x-2)^2,x]^{\top}$ and
\begin{equation}	V(x_1;\tilde{\textbf{y}}_1,\tilde{\textbf{g}}_1)=\left[y_1-(x_2^{(i)})^2-x_1^{2}\right]^2+\left[y_2-(x_2^{(i)}-2)^2-(x_1-2)^{2}\right]^2+x_1^2,
\end{equation}
while the pdf $p(x_2|\textbf{y},x_1^{(i)})\propto p(\textbf{y}|x_2,x_1^{(i)})p(x_2)$ can be expressed as $p(x_2|\textbf{y},x_1^{(i)})\propto \exp\{-V(x_2;\tilde{\textbf{y}}_2,\tilde{\textbf{g}}_2)\}$ where $\tilde{\textbf{y}}_1=[y_1-(x_1^{(i)})^2,y_2-(x_1^{(i)}-2)^2,0]^{\top}$, $\tilde{\textbf{g}}_2(x)=[x^2,(x-2)^2,x]^{\top}$ and
\begin{equation}	V(x_2;\tilde{\textbf{y}}_2,\tilde{\textbf{g}}_2)=\left[y_1-(x_1^{(i)})^2-x_2^{2}\right]^2+\left[y_2-(x_1^{(i)}-2)^2-(x_2-2)^{2}\right]^2+x_2^2.
\end{equation}
Since the marginal potentials and the nonlinearities are convex, we can use the GARS technique to sample the conditional pdf's. 

We have generated $N=10,000$ samples from the Markov chain, with fixed observations $y_1=5$ and $y_2=2$. The average acceptance rate of the GARS algorithm was $\approx 30\%$ both for $p(x_1|\textbf{y},x_2)$ and $p(x_2|\textbf{y},x_1)$. Note that this rate is indeed as a average because, at each step of the chain, the target pdf's are different (if, e.g., $x_1^{(i)}\neq x_1^{(i-1)}$ then $p(x_2|\textbf{y},x_1^{(i)})\neq p(x_2|\textbf{y},x_1^{(i-1)})$). 

\begin{figure*}[htb]
\centering
\centerline{
		\includegraphics[width=5cm,height=4cm]{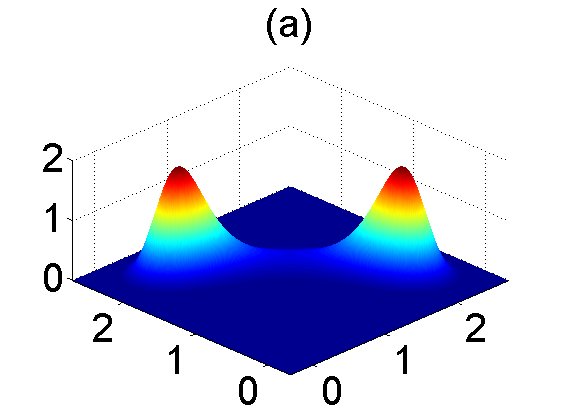}
 		\includegraphics[width=5cm,height=4cm]{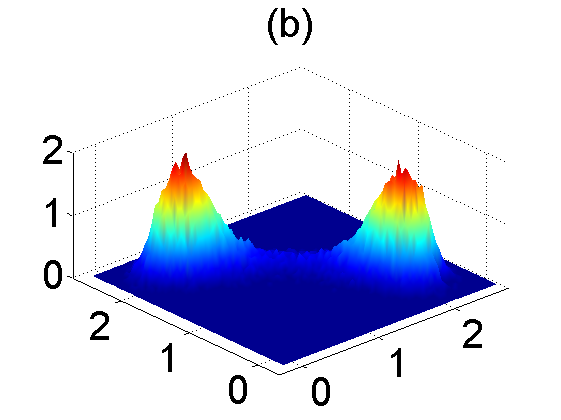}
 		}
\caption{(a) The target density $p(\textbf{x}|\textbf{y})=p(x_1,x_2|\textbf{y})\propto p(\textbf{y}|x_1,x_2)p(x_1)p(x_2)$. (b) The normalized histogram with $N=10,000$ samples, using the GARS algorithm within a Gibbs sampler.}
\label{figejemplo4}
\end{figure*}

Figure \ref{figejemplo4} (a) shows the shape of the true target density $p(x_1,x_2|\textbf{y})$, while Figure \ref{figejemplo4} (b) depicts the normalized histogram with $N=10,000$ samples. We observe that it approximates closely the shape of target pdf. 

Finally, it is illustrative to consider the computational savings attained by using the GARS method when compared with a rejection sampler with a fixed bound. Specifically, we have run again the Gibbs sampler to generate a chain of $10,000$ samples but, when drawing from $p(x_1|\textbf{y},x_2)$ and $p(x_2|\textbf{y},x_1)$, we have used RS with prior proposals ($p(x_1)$ and $p(x_2)$, respectively) and a fixed bound computed (analytically) with the method in Section \ref{sectgamma} for quadratic potentials. The average acceptance rate for the rejection sampler was $\approx 4\%$ and the time needed to generate the chain was approximately $10$ times the time needed in the simulation with the GARS algorithm. 

\section{Conclusions} \label{sConclusions}

We have proposed families of generalized rejection sampling schemes that are particularly, but not only, useful for efficiently drawing independent samples from {\it a posteriori} probability distributions. The problem of drawing from posterior distributions appears very often in signal processing, e.g., see the target localization example in this paper or virtually any application that involves the estimation of a physical magnitude given a set of observations collected by a sensor network. We have introduced two classes of schemes. The procedures in the first class are aimed at the computation of upper bounds for the likelihood function of the signal of interest given the set of available observations. They provide the means to (quickly and easily) design sampling schemes for posterior densities using the prior pdf as a proposal function. Then, we have elaborated on the bound-calculation procedures to devise a generalized adaptive rejection sampling (GARS) algorithm. The latter is a method to construct a sequence of proposal pdf's that converge towards the target density and, therefore, can attain very high acceptance rates. It should be noted that the method introduced in this paper includes the classical adaptive rejection sampling scheme of \cite{Gilks92} as a particular case. We have provided some simple numerical examples to illustrate the use of the proposed techniques, including sampling from multimodal distributions (both with fixed and adaptive proposal functions) and an example of target localization using range measurements. The latter problem is often encountered in positioning applications of sensor networks.     

\section{Acknowledgements}

This work has been partially supported by the Ministry of Science and Innovation of Spain (project MONIN, ref. TEC-2006-13514-C02-01/TCM, and program Consolider-Ingenio 2010, project CSD2008-00010 COMONSENS) and the Autonomous Community of Madrid (project PROMULTIDIS-CM, ref. S-0505/TIC/0233).

\section*{Appendix}
\label{appendix}

\noindent \textbf{Proposition}: The state estimators $\hat{x}_{j}\in \arg\max\limits_{x\in[{\mathcal B}_j]}{\ell(x|\textbf{y},\textbf{g})}=\arg\min\limits_{x\in[{\mathcal B}_j]}{V(x;\textbf{y},\textbf{g})}$ belong to the interval $\mathcal{I}_{j}$, i.e., 
\begin{equation}
\hat{x}_{j}\in \mathcal{I}_{j}\dfn [\min{(\mathcal{X}_{j})},\max{(\mathcal{X}_{j})}],	
\end{equation}
where $\mathcal{X}_{j}\dfn\{x_{1,j},\ldots,x_{n,j}\}$ is the set of all simple estimates in $\mathcal{B}_{j}$ and $\mathcal{I}_{j}\subseteq \mathcal{B}_{j}$. 

\noindent \textbf{Proof}: We have to prove that the derivative of the system potential function is
\begin{equation}
\frac{dV}{dx}<0, \ \ \mbox{for all} \ \ x<\min{(\mathcal{X}_{j})} \quad (x \in [{\mathcal B}_j]), 
\end{equation}
and 
\begin{equation}
\frac{dV}{dx}>0, \ \ \mbox{for all} \ \ x>\max{(\mathcal{X}_{j})} \quad (x \in [{\mathcal B}_j]), 
\end{equation}
so that all stationary points of $V$ stay inside $\mathcal{I}_{j}=[\min{(\mathcal{X}_{j})} ,\max{(\mathcal{X}_{j})}]$. Routine calculations yield the derivative 
\begin{equation}
\frac{d V}{d x}=-\sum^{n}_{i=1} \frac{d g_{i}}{dx}  \left[\frac{d \bar{V}_{i}}{d \vartheta_{i}}\right] _{\vartheta_i=y_{i}-g_{i}(x)}
\label{eqKK}
\end{equation}
and we aim to evaluate it outside the interval $\mathcal{I}_{j}$. To do it, let us denote $x_{min}=\min(\mathcal{X}_{j})$ and $x_{max}=\max(\mathcal{X}_{j})$ and consider the cases $\frac{dg_i}{dx}>0$  and $\frac{dg_i}{dx}<0$ separately (recall that we have assumed the sign of $\frac{dg_i}{dx}$ to remain constant in $\mathcal{B}_{j}$). 

When $\frac{dg_i}{dx}>0$ and since, for every simple estimate, $x_{i,j}\geq x_{min}$, we obtain that $y_{i}=g_{i}(x_{i,j})\geq g_{i}(x_{min})>g_{i}(x)$ $\forall x< x_{min}$. Then $y_i-g_{i}(x)>0$, for all $x<x_{min}$, and, due to properties (P1) and (P2) of marginal potential functions, $\left[\frac{d \bar{V}_{i}}{d \vartheta_{i}}\right]_{\vartheta_i=y_{i}-g_{i}(x) > 0 } > 0$ for all $i$.  As a consequence, $\frac{dV}{dx}<0$ $\forall x<x_{min}$, $x \in [{\mathcal B}_j]$. 

When $\frac{dg_i}{dx}<0$ and $x_{i,j}\geq x_{min}$, we obtain that $y_{i}=g_{i}(x_{i,j})\leq g_{i}(x_{min})<g_{i}(x)$, $\forall x<x_{min}$. Then $y_{i}-g_{i}(x)<0$ for all $x<x_{min}$ and $\left[\frac{d \bar{V}_{i}}{d \vartheta_{i}}\right]_{\vartheta_i=y_{i}-g_{i}(x) < 0 } < 0$, again because of (P1) and (P2). As a consequence, $\frac{dV}{dx}<0$ $\forall x<x_{min}$, $x \in [{\mathcal B}_j]$. 

A similar argument for $x>x_{max}$ yields $\frac{dV}{dx}>0$ for all $x>x_{max}$ and completes the proof. $\Box$

\bibliographystyle{IEEEbib}
\bibliography{bibliografia} 
   
\end{document}